\documentclass[11pt,a4paper]{article}
\pdfoutput=1
\usepackage{color}
\usepackage{bm}
\usepackage{amsmath,amssymb}
\usepackage{graphicx}

\numberwithin{figure}{section}
\numberwithin{equation}{section}

\usepackage{hyperref} 
\usepackage{cite}
\hypersetup{colorlinks=true, citecolor=green, filecolor=black, linkcolor=blue, urlcolor=blue }

\newcommand\ie{\textit{i.e.}\ }
\newcommand\eg{\textit{e.g.}\ }
\newcommand\cf{\textit{cf.}\ }

\newcommand{\aka}{{a.k.a.}\ }
\newcommand{\etc}{{\it etc.}\ }

\newcommand{\half}{\frac{1}{2}}

\newcommand{\be}{\begin{equation}}
\newcommand{\ee}{\end{equation}}
\newcommand{\bea}{\begin{eqnarray}}
\newcommand{\eea}{\end{eqnarray}}

\newcommand{\dbar}{\mathchar'26\mkern-11mu\mathrm{d}}
\newcommand{\delbar}{\mathchar'26\mkern-11mu\mathrm{\delta}}
\newcommand{\cutarg}{(-\nabla^2)}
\newcommand{\pcutarg}{\left( \frac{p^2}{\Lambda^2} \right)}
\newcommand{\KoL}{\dot{\Delta}}
\newcommand{\tr}{{\rm tr}}

\newcommand{\Sv}{\mathcal{S}}
\newcommand{\Dm}{\mathcal{D}}
\newcommand{\cL}{\mathcal{L}}
\newcommand{\cO}{\mathcal{O}}
\newcommand{\ca}{\mathsf{a}}

\newcommand{\com}[1]{}
\newcommand{\new}[1]{#1}
\newcommand{\old}[1]{}

\begin{document}

\begin{titlepage}
\begin{center}
{\huge \bf Manifestly diffeomorphism invariant classical Exact Renormalization Group}
\end{center}

\begin{center}
{\bf Tim R. Morris and Anthony W. H. Preston}
\end{center}

\begin{center}
{\it {Department} of Physics and Astronomy,  University of Southampton\\
Highfield, Southampton, SO17 1BJ, U.K.}\\
\vspace*{0.3cm}
{\tt  T.R.Morris@soton.ac.uk, awhp1g12@soton.ac.uk}
\end{center}
\abstract{We construct a manifestly diffeomorphism invariant Wilsonian (Exact) Renormalization Group for classical gravity, and begin the construction for quantum gravity.
We demonstrate that the effective action can be computed without gauge fixing the diffeomorphism invariance, and also without introducing 
a background space-time.
We compute classical contributions both within a background-independent framework and by perturbing around a fixed background, and verify that the results are equivalent.  We derive the exact Ward identities for  action{s} and kernel{s} and verify consistency.
We formulate two forms of the flow equation corresponding to the two choices of classical fixed-point: the Gaussian fixed point, and the scale invariant interacting fixed point using curvature-squared terms.
We suggest how this programme may completed to a fully quantum construction.
} 
\end{titlepage}

\tableofcontents

\section{Introduction}

In this paper, we develop a manifestly diffeomorphism invariant Wilsonian exact RG (Renormalization Group) for classical gravity. Such a construction involves a diffeomorphism invariant generalisation of a momentum cutoff $\Lambda$, allowing short distance modes with wavelength $\lesssim 1/\Lambda$ to be `integrated out' exactly (in a manner that will be made precise later) while respecting diffeomorphism invariance at all stages, resulting in an effective action $S$ that incorporates these short-distance fluctuations. $S$ can then be used as an exact alternative action to describe the dynamics of gravity on distance scales larger than $1/\Lambda$. Even at the classical level such a construct may be important, for example applied to ``cosmological back-reaction'' (see \eg  \cite{Buchert:1999er,Rasanen:2003fy,Preston:2014tua,Evans:2015zwa}, and since the transformation is exact may help settle some recent controversy \cite{Buchert:2015iva,Green:2015bma,Visser:2015mur}).
However our main motivation is that this is a stepping stone to a fully quantum manifestly diffeomorphism invariant exact RG for use in quantum gravity.
On the one hand the renormalization group structure of quantum gravity is surely of importance \cite{Stelle:1976gc,Adler:1982ri,Weinberg:1980,Reuter:1996}
and on the other hand one would hope that conceptual and computational advances would result from a  framework 
which allows computations to be done while
keeping exact diffeomorphism invariance at every stage, \ie without gauge fixing. Furthermore, as we will see, the framework allows these computations to be done without first choosing the space-time manifold and in particular without introducing a separate background metric dependence.

The framework we propose is an adaptation of the methods developed in gauge theories over a number of years, which allow continuum computations without fixing the gauge. This is achieved by utilising the freedom to design manifestly gauge invariant versions of the continuum realisation of Wilson's renormalization group (christened exact RG in ref. \cite{Wilson:1973}). Such manifest gauge invariance was first incorporated into the exact RG in ref. \cite{Morris:1995he}, however in the limited context of pure $U(1)$ gauge theory.  Following ref. \cite{Morris:1998kz} it was generalised and extensively studied first for $SU(N)$ Yang-Mills theory, then QCD \cite{Morris:2006in} and QED \cite{Arnone:2005vd,Rosten:2008zp}. For these gauge theories, regularisation is based on gauge-invariant higher derivatives supplemented by gauge invariant Pauli-Villars fields \cite{Morris:1999px}, which it was later realised could be simply understood as arising from a spontaneous breakdown of an $SU(N|N)$ super-Yang-Mills theory \cite{Morris:2000fs,Morris:2000jj}. The regularisation structure was separately studied in refs. \cite{Arnone:2000bv, Arnone:2000qd,Arnone:2001iy} and proven to work to all orders in perturbation theory. The computational methods were generalised in refs. \cite{Arnone:2002yh,Arnone:2002qi,Arnone:2002fa,Arnone:2003pa,Arnone:2002cs} so that universal results could be extracted in a way which was manifestly independent of the detailed form of the regularisation structure, and such that general group invariants could be handled \cite{Arnone:2005fb}. Using these techniques, the initial computation of the one-loop $\beta$ function at infinite $N$ \cite{Morris:1998kz} was generalised to finite $N$ \cite{Arnone:2002qi,Arnone:2002fb,Gatti:2002kc,Arnone:2002cs}, then to two loops \cite{Morris:2005tv,Rosten:2004aw,Arnone:2005fb,Rosten:2005qs,Rosten:2005ka}, extended to all loops in refs. \cite{Rosten:2005ep,Rosten:2006tk} and to computation of gauge invariant operators in refs. \cite{Rosten:2006qx,Rosten:2006pd}. For reviews and further advances see refs. \cite{Arnone:2006ie,Rosten:2010vm,Rosten:2011ty}.

%
%

When these ideas are applied to gravity a further advantage of the formalism is immediate. In continuum approaches to quantum gravity, the first step has been to express the full metric $g_{\mu\nu}$ in terms of a background metric $\bar{g}_{\mu\nu}$ in a fixed coordinate system (for example flat $\bar{g}_{\mu\nu}=\delta_{\mu\nu}$) and fluctuations $h_{\mu\nu}$ about this, essentially so that a propagator can be defined for $h_{\mu\nu}$ after appropriate gauge fixing. This means that\new{,} from the beginning\new{,} the formulation actually depends on two metrics $g_{\mu\nu}$ and $\bar{g}_{\mu\nu}$. Extra conditions are then required in order to ensure that ultimately results are \emph{background-independent}. But these can be difficult to implement exactly 
and may be too restrictive (for example requiring $h_{\mu\nu}$ to be on shell to obtain background-independence through gauge fixing independence).\footnote{From the asymptotic safety literature see for example refs. \cite{Reuter:2009kq,Becker:2014qya,Dietz:2015owa}.}  
Since\new{,} in the manifestly gauge invariant exact RG\new{,} the r\^ole of the propagator is played by a gauge invariant kernel whose form is part of the freedom allowed in designing the Kadanoff blocking, the problem of inverting a propagator does not arise. As we will see this allows computations to be done entirely in terms of the full quantum metric $g_{\mu\nu}$. In this way a background metric $\bar{g}_{\mu\nu}$ is never introduced and the issue of background independence thus never arises.

In fact, since the flow equation is designed to ensure that the Wilsonian action remains quasi-local, \ie such that the effective Lagrangian can be expanded in powers of space-time derivatives, we will see that (to any finite order in this expansion) it is not necessary to make any {\it a priori} assumptions about the space-time manifold (beyond its smoothness). The entire computation can be phrased in terms of manipulations of covariant derivatives.
The resulting Lagrangian can be computed iteratively in terms of (covariant derivatives of) curvature invariants of increasing dimension,  as we will see explicitly in this paper at the classical level.

Nevertheless, more insight can be gained by organising the result as an expansion in $n$-point vertices for fluctuations $h_{\mu\nu}$ about a particular background. As an example, we choose $\bar{g}_{\mu\nu}=\delta_{\mu\nu}$ and show that in this way the full momentum dependence of the $n$-point vertices can be computed iteratively about this background (\ie  from the already-solved $m<n$ point vertices). It will be clear that the same Lagrangian is being computed in these alternative approaches, however we also confirm this through some consistency checks. When expanded in fluctuations in this way the diffeomorphism invariance is obscured, but is nevertheless present and verified through exact Ward identities that we also derive.

As we will recall, an infinitessimal step in the flow of  the exact RG is just an exact change of field variables. At the quantum level, the partition function is unchanged by the exact RG. At the classical level the effective action satisfies the same equations of motion as the original (bare) action, albeit now in terms of effective field variables. 

The current paper is limited to classical computations. If we were to attempt quantum corrections with the current set-up we would find ultraviolet divergences. A research direction for furnishing the extra structure necessary to provide full regularisation is described in sec. \ref{sec:conclusions}.

One aspect of a fully quantum flow is necessarily anticipated in the structure of the flow equation itself. Yang-Mills theory (in four space-time dimensions) has a well defined continuum limit given by constructing the theory around the Gaussian fixed point (\ie with vanishing Yang-Mills coupling). Therefore \old{it is clear that} the flow equation should be adapted for use around this fixed point, as was done in our earlier papers. Gravity as described with the Einstein-Hilbert action is not perturbatively renormalisable as a quantum theory, meaning in Wilsonian language that Newton's constant $G$ is irrelevant, and that the continuum limit results in non-interacting (linearised) gravitons. Nevertheless much can be learned from the effective field theory description organised in terms of increasing powers of $G\sim 1/M^2_{\rm Planck}$ \cite{Donoghue:1994dn}, therefore in this paper we construct a manifestly diffeomorphism invariant flow equation that naturally develops such an expansion while allowing for the fact that $G$ becomes a running coupling in general.

If an asymptotically safe fixed point exists \cite{Weinberg:1980,Reuter:1996,Reuter:2012} both the classical and quantum parts of the flow equation would be equally important. Since we keep only the classical part, it is not entirely clear what the best adapted structure for the flow equation is in this case. Instead we supply a form of the flow equation adapted to the renormalisable  $O(\partial^4)$ gravity as developed in refs. \cite{Stelle:1976gc,Adler:1982ri}, which has problems with unitarity, but which might reasonably be expected to be a closer classical analogue.

The paper is structured as follows. In the next section we review the elements of the construction of manifestly gauge invariant flow equations that we will need, and then in sec. \ref{sec:biflow} adapt these to the construction of a diffeomorphism invariant and background-independent flow equation for gravity.  We  see that as well as introducing the differentiated ``effective propagator'' $\dot{\Delta}(-\nabla^2)$ part of the kernel, which we choose to take the simple covariantisation indicated, we need to introduce two trace structures and a corresponding DeWitt parameter $j$. We point out the special cases associated \new{with} conformally reduced gravity and unimodular gravity. In sec. \ref{sec:dimensions} we use dimensional analysis firstly in $D$ dimensions for further insight into why the (now dimensionful) gauge coupling $g$ appears as discussed in sec. \ref{sec:gauge}. We then to adapt this to discuss the r\^ole of Newton's coupling (and the cosmological constant) in the gravity case and their relation to couplings in the effective action. We also provide a first discussion of the two schemes: the Weyl scheme and the Einstein scheme, and constrain the form of $\Delta$ in these two schemes. This in turn leads us to introducing classical Lagrangians with dimension $\ell=4$ and $2$ respectively. In sec. \ref{sec:bieffact} we describe in general terms how the classical effective action can then be computed iteratively as an expansion in local diffeomorphism invariant scalar operators $\cO_d$. In secs. \ref{sec:bifixed}  and \ref{sec:biEHfixed}
we then apply this to the computation of the fixed point effective action and determination of the seed action\old{,} in the Weyl and Einstein schemes. In both schemes renormalisation conditions are required to define them precisely;  these fix certain couplings.
In secs. \ref{sec:biaway} and \ref{sec:biEHfixed} we also point out the relevant operators at the classical level, and compute the exact classical flow equations for these. The seed action is determined by the requirement that it and the fixed point effective action have the same two-point vertex when expanded around a flat background. This also determines the form of the kernel and fixes the value of $j$ (up to a further discrete choice in the Einstein case). In sec. \ref{sec:fbflow} we introduce the expansion around flat background, which means that the actions are best expressed in terms of $n$-point vertices in momentum space. Exact diffeomorphism invariance still governs the equations but through exact Ward identities, which we derive for the action in sec. \ref{sec:WIa} and for the kernel in \ref{sec:WIk}. We provide a consistency check on these equations in sec. \ref{sec:check}, and in sec. \ref{sec:kernelpoint} demonstrate how to compute the $n$-point kernel vertices. In sec. \ref{sec:WIa} we also derive the differential Ward identities to demonstrate that the classical effective action can be computed iteratively in terms of increasingly higher $n$-point vertices, and to demonstrate that the two-point vertex splits into a momentum independent cosmological constant part (\new{with} support by the expansion in app. \ref{sec:sqrtg})
and a transverse part. In sec. \ref{sec:transverse} we provide  the linearly independent transverse two-point momentum structures, and finally these are put to use in sec. \ref{sec:2pts} to compute the form of the classical fixed-point two-point  vertices in the two schemes, thus finally determining also the effective propagator and the simplest form for the seed action in the two schemes. Sec. \ref{sec:conclusions} further discusses the construction, in particular where there are choices and where there are not (with support from app. \ref{app:seed0}), and outlines a possible route to a fully quantum manifestly diffeomorphism invariant exact RG.

\section{Mini review of manifestly gauge invariant exact RGs}
\label{sec:review}

In this section we review the main ideas that we will need to adapt for construction of  a manifestly diffeomorphism invariant exact RG.

\subsection{Kadanoff blockings}
\label{sec:kadanoff}

We begin the derivation of the exact RG with a Kadanoff blocking procedure \cite{Kadanoff:1966wm}.
Kadanoff blockings are averaging schemes used to infer the macroscopic behaviour of a system from its microscopic physics.
The original formulation envisages a large lattice of spins.
The method assumes that correlations between spins can be completely attributed to interactions between close neighbours.
This notion of locality is an essential feature of the scheme.

The blocking scheme groups lattice sites into blocks, with each block averaged to a single spin state.
In general, these spins have different interaction strengths with their neighbours than the original, microscopic spins do with theirs.
There are an infinite number of different Kadanoff blockings, and so in turn there are an infinite number of different Wilsonian RGs \cite{Latorre:2000qc}.

Adapting the method from statistical mechanics to field theory requires a continuum definition of Kadanoff blockings for continuous fields \cite{Wilson:1971bg,Wilson:1971dh}.
Instead of averaging blocks of spins, one integrates out momentum modes down to some smooth cutoff set by some Lorentz invariant momentum scale, $\Lambda$ \cite{Wegner:1972ih,Wilson:1973,Wilson:1974mb}.
Actually, an immediate requirement to maintain a notion of locality is that the metric should be rotated into a Euclidean signature.
This is because light-like separations in a Lorentzian metric can have arbitrarily large coordinate separations for a zero invariant interval.

Consider an effective (\ie macroscopic) scalar field $\varphi$ whose physics is described by an effective action $S[\varphi]$.
Given a bare (\ie microscopic) field $\varphi_{0}$ and a bare action $S_{\rm bare}[\varphi_0]$, the standard definition for a Kadanoff blocking is via
\begin{equation}\label{Kblocking}
 e^{-S[\varphi]}=\int\Dm\varphi_{0} \ \delta\left[\varphi-b\left[\varphi_{0}\right]\right]e^{-S_{\rm bare}\left[\varphi_{0}\right]}.
\end{equation}
The blocking functional is, in turn, a scalar field with a position argument. A simple linear example of a blocking functional in a $D$-dimensional field theory is
\begin{equation}
 b[\varphi_{0}](x) = \int_{y} B(x-y)\varphi_{0}(y),
\end{equation}
where $B(z)$ is a kernel that provides a smooth infrared cutoff such that $B(z)$ decays rapidly towards zero once $|z|\Lambda>1$.
This allows us to integrate out the higher momentum modes while keeping our effective action as an expansion in local operators.
We use a shortened notation for a $D$ dimensional integral over a set of spatial coordinates, $x$, such that $\int_{x}\equiv\int d^{D}x$ for convenience.

From equation (\ref{Kblocking}), we can integrate the effective Boltzmann factor over the effective field to obtain the partition function.
On the right hand side, because of the delta function, we can integrate out the effective field to get the same partition function we would obtain using the bare field and the bare action \ie the partition function is invariant under change of cutoff scale and the blocking procedure has not changed the physics:
\begin{equation}\label{partinv}
 \mathcal{Z} = \int\Dm\varphi\ e^{-S[\varphi]} = \int\Dm\varphi_{0}\ e^{-S_{\rm bare}[\varphi_{0}]}.
\end{equation}

To obtain an exact RG, we differentiate the effective Boltzmann factor with respect to `RG time':
\begin{equation}\label{BoltzFlow}
 \Lambda\frac{\partial}{\partial\Lambda}e^{-S[\varphi]}=-\int_{x} \frac{\delta}{\delta \varphi(x)}\int\Dm\varphi_{0}\ \delta\left[\varphi-b\left[\varphi_{0}\right]\right]\Lambda\frac{\partial b[\varphi_{0}](x)}{\partial\Lambda}e^{-S_{\rm bare}\left[\varphi_{0}\right]}
\end{equation}
In the above equation, the functional integral over the bare field thus yields some function of $x$ and functional of $\varphi$, which we write as $-\Psi(x)e^{-S[\varphi]}$, where $-\Psi(x)$ can be thought of roughly as the rate of change of the blocking functional with respect to RG time. We thus have:
\begin{equation}\label{generalERG}
 \Lambda \frac{\partial}{\partial \Lambda}e^{-S[\varphi]}=\int_{x} \frac{\delta}{\delta \varphi(x)}\left(\Psi(x)e^{-S[\varphi]}\right)\,,
\end{equation}
from which we of course obtain
\be 
\label{generalERG2}
\Lambda \frac{\partial}{\partial \Lambda} S =  \int_{x} \Psi(x)\frac{\delta S}{\delta \varphi(x)}  - \int_{x} \frac{\delta \Psi(x)}{\delta \varphi(x)}\,.
\ee
This is now the general form for constructing an exact RG flow equation for a single scalar field.
Since there are infinitely many blocking functionals, there are infinitely many possible flow equations that leave the partition function invariant under change of cutoff.
The invariance can now be seen simply by noticing that this form is a total functional derivative in $\varphi$, which can be functionally integrated with respect to $\varphi$ to give the rate of change of partition function, and which is zero for suitably well behaved Boltzmann factor. Furthermore for later purposes note that the change in the effective action $\delta S$  induced by flow from $\Lambda$ to $\Lambda-\delta\Lambda$ is just the result of the change of field variable $\varphi$ to $\varphi-\Psi\delta\Lambda/\Lambda$, the $\delta\Psi/\delta\varphi$ term coming from reparametrising the measure in \eqref{partinv}.

It will be convenient from now on to represent differentiation with respect to RG time by an over-dot such that, for some function $f(\Lambda)$, $\dot{f}(\Lambda):=\Lambda\frac{\partial}{\partial \Lambda}f(\Lambda)$.
It is also conventional to introduce the following notation, as used \eg in refs. \cite{Morris:1998,Arnone:2006ie}:
\begin{equation}\label{kernelquick}
 f\cdot W\cdot g := \int_{x} f(x) 
 W\left(-\partial^2\right)g(x),
\end{equation}
where $W$ is a (Lorentz invariant) momentum kernel,  and as we will see, usually is related to a term that can be thought of as an effective propagator at a fixed point. As such, by dimensions it can be written as a dimensionful part depending on $-\partial^2$ only, times a dimensionless function of $-\partial^2/\Lambda^2$. (To simplify notation, we will usually leave implicit the dependence of the kernel and effective action on $\Lambda$.)

\subsection{Flow equations for massless scalar fields}
\label{sec:scalars}

We now wish to specialize (\ref{generalERG}) to give us the Polchinski form for the flow equation of a scalar field \cite{Polchinski:1983gv}.
The rate of change of the blocking functional can in this case be expressed as
\begin{equation}\label{Polchange}
 \Psi(x) = \frac{1}{2}\int_{y} \dot{\Delta}(x,y)\frac{\delta \Sigma}{\delta \varphi(y)},
\end{equation}
where $\Delta=c(p^2/\Lambda^2)/p^{2}$ is indeed the effective propagator, which has been regulated with an ultraviolet cutoff function, $c(p^2/\Lambda^2)$, and $\Sigma$ is in the form of an action.
More specifically, $\Sigma = S - 2\hat{S}$, where $\hat{S}$ is a functional of fixed form, called the `seed action', and is an action whose only scale is $\Lambda$. 
There is a great deal of freedom in the choice of the seed action, 
without changing the underlying physics. This is part of the freedom of choice of how we implement Kadanoff blocking. As we will see shortly, the required notion of locality in this context is implemented by insisting that $c(p^2/\Lambda^2)$ has a Taylor expansion to all orders and that $\hat{S}$ similarly has a derivative expansion to all orders, \ie is quasi-local \cite{Morris:1999px}.
As will become apparent, a useful choice for $\hat{S}$ is simply the regularized kinetic term in the effective action; it is given in position representation by
\begin{equation}
\label{scalarSeed}
 \hat{S} = \frac{1}{2} \partial_{\mu}\varphi\cdot c^{-1}\cdot\partial_{\mu}\varphi,
\end{equation}
where we use the notation introduced in (\ref{kernelquick}) and there is an implicit summation over the index, $\mu$, remembering that the metric has been rotated into Euclidean signature. 
This choice of seed action leads us to the Polchinski form of the flow equation.
However, we can add further 3-point and higher corrections to this seed action without altering the continuum physics \cite{Arnone:2002yh,Arnone:2003pa,Arnone:2006ie}. 

For canonical normalisation of the effective propagator and the kinetic terms \eqref{scalarSeed}, we require $c(0)=1$. Actually, as we will see, requiring that we can canonically normalise simultaneously both the kinetic terms and $\Delta$, determines the factor of half in \eqref{Polchange}; saying it differently the integrated kernel turns out to be normalised as $1/2p^2$ for small $p$, which we then express as $\half \Delta$ so that $\Delta$ has the canonical normalisation of the propagator. This observation will be useful later for the gravity flow equation.

Substituting (\ref{Polchange}) into (\ref{generalERG}), we obtain the flow of the action in position representation with respect to RG time:
\begin{equation}\label{scalarSdot}
 \dot{S} = \frac{1}{2} \frac{\delta S}{\delta\varphi}\cdot\dot{\Delta}\cdot\frac{\delta \Sigma}{\delta\varphi} - \frac{1}{2} \frac{\delta}{\delta \varphi}\cdot\dot{\Delta}\cdot\frac{\delta \Sigma}{\delta\varphi}.
\end{equation}
Since $\dot{\Delta} = -2c'(p^2/\Lambda^2)/\Lambda^2$ has a Taylor expansion and $\hat{S}$ has a derivative expansion, we see that an effective action $S$ that is quasi-local to begin with{,} remains quasi-local under the flow for any finite RG time \cite{Morris:1999px}.

\begin{figure}[ht]
\begin{center}
\includegraphics[height=1.8cm]{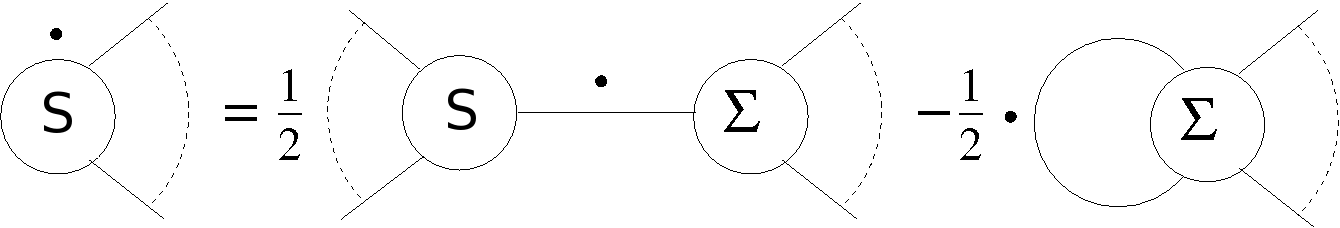}
\end{center}
\caption{Diagrammatic illustration of a generalised flow equation for scalar fields}
\label{fig:npointflow}
\end{figure}

One obtains the flow equations for $n$-point functions from this by taking $n$ functional derivatives with respect to $\varphi$ of both sides and taking the $\varphi\to 0$ limit.
This can be illustrated digrammatically as in Figure \ref{fig:npointflow}, adapted from \cite{Arnone:2006ie}.

In Figure \ref{fig:npointflow}, $n$-point functions are represented by solid circles labelled with the name of the action inside, effective propagators and external legs are represented by solid lines coming out of the $n$-point functions, and the over-dots represent differentiation of individual actions and propagators with respect to RG time.
An advantage in this diagrammatic view is that it provides an intuitive picture of the flow equation.
For example, while the first term on the right hand side has a classical part, it is clear that the second term has no classical part, since there is a propagator linked to $\Sigma$ at both ends, forming a loop.
Thus the tree-level part of the RG flow equation does not require the second term.

Let us now consider the 2-point part of the tree-level flow equation in momentum representation for a single component scalar field theory invariant under $\varphi\to -\varphi$.
Since this scalar theory has no 1-point functions 
and both the action and the seed action have the same 2-point function, the tree-level 2-point flow equation is easily expressed only in terms of the tree-level 2-point function, denoted here by ${S}^{\varphi\varphi}$,
\begin{equation}
\label{scalar-2pt-flow}
 \dot{S}^{\varphi\varphi} = -S^{\varphi\varphi}\dot{\Delta}S^{\varphi\varphi}.
\end{equation}
We see that this is consistent with the choice $\Delta=\left(S^{\varphi\varphi}\right)^{-1}$ that we already made. Later we will use such an equation to determine the form of $\Delta$ given the form of the effective two-point vertex.
Since $\Delta$ inverts the 2-point function, it can be identified as the effective propagator.
For massless scalar fields, the classical 2-point function comes purely from the kinetic term, which is the same in $S$ as in $\hat{S}$.
Higher-point modifications to $\hat{S}$ do not impact on the 2-point function at the classical level and indeed do not affect any physics at the classical or quantum level, as has been checked explicitly in \cite{Arnone:2002yh,Arnone:2003pa}.
This is because these modifications are nothing more than reparametrizations of the field as the high energy modes are integrated out \cite{Latorre:2000qc}.

Since we are working with dimensionful quantities, a fixed point action is characterised by the fact that the only scale appearing in it is $\Lambda$. To see this, note that if we had also performed the rescaling step part of the Wilsonian RG by redefining all dimensionful quantities to be dimensionless, using the appropriate power of $\Lambda$, the action would then no longer contain any functional dependence on $\Lambda$. In other words it would indeed be a fixed point action under evolution in $\Lambda$.

By choosing $\hat{S}$ to be only the kinetic term as in \eqref{scalarSeed}, we have enabled a closed solution $S=\hat{S}$ for the fixed point action. According to the standard Wilsonian construction, the continuum limit is then constructed by adding relevant perturbations to this (see for example ref.  \cite{Morris:1998}). \old{Of course} \new{O}ne then discovers the infamous triviality problem, namely that all interactions are either irrelevant or marginally irrelevant. However nevertheless it is useful to work with the effective theory with a marginally irrelevant four-point coupling (the Higgs sector of the Standard Model being just one example).

\subsection{Application to Yang-Mills theories}\label{sec:gauge}

Now let us put aside the scalar field $\varphi$ and consider a gauge field $A_{\mu}$ (valued in some Lie algebra).
Manifest gauge invariance requires that the connection can have no wavefunction renormalization.
The gauge field itself inherits this property if the covariant derivative is defined as:
\begin{equation}
\label{gaugeCovariantDerivative}
 D_\mu := \partial_\mu -iA_\mu.
\end{equation}
To see that we require there to be no wavefunction renormalization, note that the gauge transformation of the field is
\begin{equation}
 \delta A_\mu = [D_\mu,\omega(x)].
\end{equation}
Changing our variable to a renormalized field, $A^R_\mu = Z^{-1/2}A_\mu$, the transformation becomes
\begin{equation}
 \delta A^{R}_\mu = Z^{-1/2}\partial_\mu \omega-i[A^{R}_{\mu},\omega].
\end{equation}
Thus gauge invariance is preserved only if we fix $Z=1$.
This conclusion cannot be made in the more conventional approach, which fixes a gauge, because $\omega$ is replaced by a ghost field \cite{Faddeev:1967fc,Becchi:1974xu,Becchi:1974md,Becchi:1975nq} thus the second term becomes a composite operator which requires its own renormalisation. 
The field strength is $F_{\mu\nu} := i[D_\mu,D_\nu]$.
The action is written in a form where the coupling is seen as an overall scaling factor:
\begin{equation}\label{rescaleact}
 S[A](g) = \frac{1}{4g^2}{\rm tr}\int_{x} F_{\mu\nu}\,c^{-1}\!\left(-\frac{D^2}{\Lambda^2}\right)\!F_{\mu\nu} + \mathcal{O}(A^3) + \cdots
\end{equation}
We have organised the expansion in terms of the minimum number of fields. Without loss of generality, we can write the higher-covariant derivative terms in the $\mathcal{O}(A^2)$ term as above, defining what we mean by the cutoff profile $c$. Note that quasi-locality then requires that $c$ is Taylor expandable and $c(0)\ne0$. In fact it is natural to insist $c(0)=1$ again, this time as the renormalisation condition to define $g$. The $g$ expansion is covered in more detail in the literature highlighted in the introduction, and it and the analogous issues for gravity will be also be discussed in more detail later in secs. \ref{sec:dimensions} and \ref{sec:biEHfixed}.
Notice also that,  like in massless scalar field theory, only the regularized kinetic term then contributes to the 2-point function.
The effective action can be expanded out loopwise as
\begin{equation}\label{rescaleactseries}
 S = \frac{1}{g^2}S_0 + S_1 + g^2 S_2 + \cdots
\end{equation}
where $S_i$ is the contribution at the $i$-loop level and the factors of $g^2$ also count powers of $\hbar$.
Similarly, the $\beta$ functions can be written as the following loopwise expansion:
\begin{equation}\label{rescalebeta}
 \beta := \Lambda\partial_{\Lambda}g = \beta_1 g^3 + \beta_2 g^5 + \cdots
\end{equation}

\begin{figure}[ht]
\begin{center}
\includegraphics[height=1.8cm]{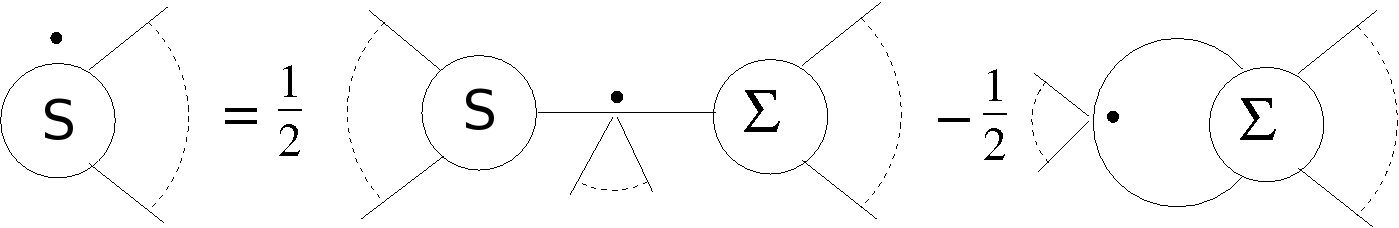}
\end{center}
\caption{Diagrammatic illustration of a gauge invariant flow equation for gauge fields}
\label{fig:Anpoint}
\end{figure}

We wish to ensure that our flow equation is gauge invariant, but this property would be broken by the kernel, $\dot{\Delta}(-\partial^2)$.
To restore gauge invariance, we need to covariantize the kernel.
There are an infinite number of ways to do this, but a simple method is to replace the partial derivatives with covariant derivatives, modifying the kernel to $\dot{\Delta}(-D^2)$.
For some choice of covariantization, we can write a gauge invariant flow equation as
\begin{equation}\label{gaugeflow}
 \dot{S} = \frac{1}{2}\frac{\delta S}{\delta A_\mu}\cdot \{\dot{\Delta}\} \cdot \frac{\delta \Sigma_g}{\delta A_\mu} - \frac{1}{2} \frac{\delta}{\delta A_\mu}\cdot\{\dot{\Delta}\}\cdot\frac{\delta \Sigma_g}{\delta A_\mu}.
\end{equation}
where we use the notation in (\ref{kernelquick}), except that the braces indicate some fixed choice for how to covariantize the kernel, and as a consequence of scaling out the coupling as in \eqref{rescaleact}, we now have $\Sigma$ replaced by
\be 
\label{Sigma-g}
\Sigma_g =  g^2 S - 2\hat{S}\,.
\ee
Covariantizing the kernel has introduced a series expansion in the field into the kernel and thus the kernel has non-zero functional derivatives with respect to the field.
Diagrammatically, this means that the kernel can now have external legs.
This property is now important when calculating the $n$-point functions of $\dot{S}$.
This can be illustrated digrammatically as in Figure \ref{fig:Anpoint}, adapted from \cite{Arnone:2006ie}.

Again, the classical level only uses the first term in \eqref{gaugeflow}, which corresponds to the first diagram on the right hand side of fig. \ref{fig:Anpoint}. Since $g^2$ now also counts $\hbar$, equivalently it can be obtained by taking the $g\to0$ limit. At the risk of causing some confusion, we now drop the subscript $0$ on the classical action, thus returning to $\Sigma \equiv   S - 2\hat{S}$ notation as in the scalar case, but retain only the first term in \eqref{gaugeflow}.
As before, we set the 2-point part of $\hat{S}$ equal to the 2-point part of $S$ at the classical level. It can be written in momentum representation as
\begin{equation}
\label{gauge-2point}
 S^{AA}_{\mu\nu} = (\delta_{\mu\nu}p^2 -p_{\mu}p_{\nu})\,c^{-1}\!\left(\frac{p^2}{\Lambda^2}\right).
\end{equation}
Gauge invariance and Poincar\'{e} invariance are sufficient to force the 2-point function to take this form, which is transverse. 
The flow equation now reads 
\begin{equation}
 \dot{S}^{AA}_{\mu\nu} = -S^{AA}_{\mu\alpha}\dot{\Delta}S^{AA}_{\alpha\nu}.
\end{equation}
Knowing that $S^{AA}_{\mu\alpha}S^{AA}_{\alpha\nu} = (p^2 c^{-1})S^{AA}_{\mu\nu}$, the solution can be taken to be $\Delta = c/p^2$, as with massless scalar field theory (with the normalisation assured by the overall factor of $1/2$ in \eqref{gaugeflow}.
Unlike in scalar field theory, the gauge invariance prevents $\Delta$ from inverting the 2-point function, thus it is no longer a true effective propagator, but rather it satisfies the condition that
\begin{equation}
 \Delta S^{AA}_{\mu\nu} = \delta_{\mu\nu}-p_{\mu}p_{\nu}/p^2.
\end{equation}
Instead of having unity on the right hand side, we have the transverse projector.

Equating the two-point vertices of $S$ and $\hat{S}$ at the classical level in a theory with no 1-point functions also has the benefit that $\dot{S}$ can be determined at the classical $n$-point level entirely in terms of $(n-1)$-point and lower functions from $S$, given some $\hat{S}$ that one is essentially free to choose \cite{Arnone:2006ie}.
This happens because all contributions to the $n$-point part of $\dot{S}$ from $n$-point functions in $S$ are cancelled in the classical flow equation. 

As remarked in the scalar case, the continuum limit (and thus the quantum field theory) needs now to be constructed by adding relevant perturbations to the fixed point action. As is well known, the coupling $g$ will turn out to be marginally relevant (otherwise known as asymptotically free) and is the only relevant direction. The fixed point action is given by the formal $g\to0$ limit of \eqref{rescaleactseries}, \ie effectively by $S_0$ (now renamed $S$). Unlike in the scalar case, there is no closed solution for this fixed point action however. It has an infinite number of vertices. Since we are free to choose the seed action we can at least insist it takes a closed form, for example:
\begin{equation}
 \hat{S}[A] = \frac{1}{4}\,{\rm tr}\!\int_{x} F_{\mu\nu}\,c^{-1}\!\left(-\frac{D^2}{\Lambda^2}\right)\!F_{\mu\nu}\,.
\end{equation}

\section{Background-independent gravity flow equation}
\label{sec:biflow}

We adopt sign conventions such that the Ricci tensor $R_{\mu\nu}=R^{\alpha}_{\ \mu\alpha\nu}$ and 
\begin{equation}
R^{\mu}_{\ \nu\rho\sigma} = 2\,\partial_{[\rho}\Gamma^{\mu}_{\ \sigma]\nu} + 2 \,\Gamma^{\mu}_{\ \lambda[\rho} \Gamma^{\lambda}_{\ \sigma]\nu}\,,
\end{equation}
where the Levi-Civita connection is defined in the usual way:
\begin{equation}\label{Levi-Civita}
 \Gamma^{\mu}_{\ \nu\rho} = \frac{1}{2}g^{\mu \alpha}(\partial_{\nu} g_{\rho \alpha} + \partial_{\rho} g_{\nu \alpha} - \partial_{\alpha} g_{\nu \rho})\,.
\end{equation}
To maintain quasi-locality, we Wick rotate such that the metric, $g_{\mu\nu}$, has Euclidean signature.
In analogy to the manifestly gauge invariant exact RG for Yang-Mills, we now wish to construct a manifestly diffeomorphism invariant exact RG for gravity.
Manifest diffeomorphism invariance gives us the opportunity for studying two formalisms: one that maintains a strict background independence and one that defines our metric as a given background $\bar{g}_{\mu\nu}$ plus a perturbation $h_{\mu\nu}$.
The latter formalism has had to be used for continuum studies in quantum gravity,   since gauge-fixing requires a fixed background (and coordinates). The typical choice is $\bar{g}_{\mu\nu}=\delta_{\mu\nu}$, which we will also use. In fact for a diffeomorphism invariant exact RG, the two formalisms are straightforwardly related, as we will see. 
In this section and in sec. \ref{sec:bieffact}, we will outline the background-independent formalism and then develop the fixed-background formalism from sec. \ref{sec:fbflow} onwards.

We begin the manifestly diffeomorphism exact RG by defining a Kadanoff blocking functional, $b_{\mu\nu}[g_{0}](x)$, which is itself a covariant tensor field, via the Boltzmann factor:
\begin{equation}\label{block}
e^{-S[g]} = \int\mathcal{D}g_{0}\ \delta\left[g-b\left[g_{0}\right]\right]e^{-S_{{\rm bare}}[g_{0}]},
\end{equation}
where $g_{0\mu\nu}$ is the bare metric.\footnote{We suppress tensor indices inside functional arguments and in the functional integral for notational convenience.}
This is directly analogous to equation (\ref{Kblocking}).
As with scalar and gauge theories, \cf eqn. (\ref{partinv}), the partition function is invariant under change of cutoff.
We obtain the exact RG flow equation as done in (\ref{BoltzFlow}) by differentiating the Boltzmann factor with respect to RG time:
\begin{equation}
 \Lambda\frac{\partial}{\partial\Lambda}e^{-S[g]}=-\int_{x} \frac{\delta}{\delta g_{\mu\nu}(x)}\int\mathcal{D}g_{0}\ \delta\left[g-b\left[g_{0}\right]\right]\Lambda\frac{\partial b_{\mu\nu}(x)}{\partial\Lambda}e^{-S_{{\rm bare}}[g_{0}]},
\end{equation}
In analogy with equation (\ref{generalERG}), we thus obtain a general exact RG for gravity in terms of the rate of change of blocking functional, $\Psi_{\mu\nu}(x)$:
\begin{equation}
\label{Psimunu-flow}
 \Lambda\frac{\partial}{\partial\Lambda}e^{-S[g]}=\int_{x} \frac{\delta}{\delta g_{\mu\nu}(x)}\left(\Psi_{\mu\nu}(x)e^{-S[g]}\right).
\end{equation}
To achieve a form of exact RG flow equation applicable to gravity, analogous to (\ref{Polchange}), we now specify the form of $\Psi_{\mu\nu}$:
\begin{equation}
\label{Psimunu}
 \Psi_{\mu\nu}(x) = \half \int_{y} K_{\mu\nu\rho\sigma}(x,y)\frac{\delta\Sigma}{\delta g_{\rho\sigma}(y)},
\end{equation}
where as in the scalar and gauge field cases, we anticipate the need for a factor $1/2$ to allow canonical normalisation, and where the  kernel, $K_{\mu\nu\rho\sigma}(x,y)$, is a covariant bitensor which can be chosen to be symmetric.
The $\mu$ and $\nu$ indices of $K_{\mu\nu\rho\sigma}$ are associated with the position argument $x$ and the $\rho$ and $\sigma$ indices are associated with the position argument $y$.
Just as in the scalar or gauge theory cases, we set $\Sigma=S-2\hat{S}$, where $\hat{S}$ is the ``seed action'' that we are essentially free to choose, whose only length scale is $\Lambda$.
This gives us an adaptation of the Polchinski flow equation, which we had in (\ref{scalarSdot}) for a pure scalar theory and (\ref{gaugeflow}) for a pure gauge theory, that now applies to gravity:
\begin{equation}
\label{full-flow}
 \dot{S}=\half \int_{x}\frac{\delta S}{\delta g_{\mu\nu}(x)}\int_{y}K_{\mu\nu\rho\sigma}(x,y)\frac{\delta\Sigma}{\delta g_{\rho\sigma}(y)}-\half\int_{x}\frac{\delta}{\delta g_{\mu\nu}(x)}\int_{y}K_{\mu\nu\rho\sigma}(x,y)\frac{\delta\Sigma}{\delta g_{\rho\sigma}(y)}\,.
\end{equation}
As is the case with scalar and gauge theories, the second term has no tree-level part.
We will be focussing on the tree-level, so we will mostly neglect this term from here on.

As we noted in the introduction and below \eqref{generalERG}, the full exact RG flow just induces an exact reparametrisation of the effective action, as is again clear from \eqref{Psimunu-flow}. The physical equivalence of the effective action at different scales $\Lambda$ is then clear. Since we will be focussing on the classical evolution only, it is worth pointing out that it is also straightforward to see the equivalence directly at the classical level. Indeed, keeping only the classical part of \eqref{full-flow} means, by \eqref{Psimunu}, that
\be 
\dot{S} = \int_{x}\Psi_{\mu\nu}(x) \frac{\delta S}{\delta g_{\mu\nu}(x)}\,,
\ee
in other words 
\be 
\label{classical_equivalence}
S_{\Lambda-\delta\Lambda}[g_{\mu\nu}] = S_{\Lambda}[g_{\mu\nu}-\Psi_{\mu\nu}\,\delta\Lambda/\Lambda]\,.
\ee

Let us 
also for convenience in what follows express $K_{\mu\nu\rho\sigma}(x,y)$, as a covariant derivative operator acting on a space-time delta function, $\delta(x-y)$, allowing the integral over $y$ to be done trivially.
One respect in which gravity differs from scalar and gauge theories is that we have two possible index structures for the flow equation. Let us illustrate this with just the classical component of \eqref{full-flow}, recognising that this in turn defines the kernel $K_{\mu\nu\rho\sigma}$ and thus also the quantum part of \eqref{full-flow}.
First, we have the ``cross-contracted'' form:
\begin{equation}\label{c.c.}
\dot{S}|_{c.c.} = \half\int_{x}\frac{\delta S}{\delta g_{\mu\nu}}\frac{g_{\mu(\rho}g_{\sigma)\nu}}{\sqrt{g}}\dot{\Delta}\cutarg \frac{\delta \Sigma}{\delta g_{\rho\sigma}}.
\end{equation} 
Next we have the ``two-traces'' form:\footnote{The $1/\sqrt{g}$ is required in order to ensure an overall density of weight $-1$. The metric factors commute with the covariant derivatives and thus with the kernel $\dot{\Delta}$, so can be placed anywhere in these expressions.}
\begin{equation}\label{t.t.}
\dot{S}|_{t.t.} = \half\int_{x}\frac{\delta S}{\delta g_{\mu\nu}}\frac{g_{\mu\nu}g_{\rho\sigma}}{\sqrt{g}}\dot{\Delta}\cutarg \frac{\delta \Sigma}{\delta g_{\rho\sigma}}.
\end{equation} 
In general, we expect that the full flow equation is a linear combination of both index structures:
\begin{equation}\label{flowratio}
\dot{S} = \dot{S}|_{c.c.} + j\dot{S}|_{t.t.},
\end{equation}
where $j$ is a dimensionless parameter. In other words the kernel is set to
\begin{equation}
\label{kernel-grav}
 K_{\mu\nu\rho\sigma}(x,y) = \frac{1}{\sqrt{g}}\delta(x-y)\left(g_{\mu(\rho}g_{\sigma)\nu}+jg_{\mu\nu}g_{\rho\sigma}\right)\dot{\Delta}\cutarg
\end{equation}
(where $\nabla^2$ acts on the $y$ dependence to the right).
Note that we need only one parameter here since we can absorb an overall factor into $\dot{\Delta}$. The remaining parameter, $j$, thus distinguishes different ways of integrating out the metric. It appears for the same reason as in the DeWitt supermetric \cite{DeWitt:1967yk}, where it is part of the apparent freedom in choice of quantization, however we will see that in the present context the other constraints we place on the form of the flow equation will determine its value.


To see how the value of $j$ affects the balance of modes propagating in the flow equation, let us briefly consider two special cases.
Firstly, a value of $j\to \infty$ corresponds to only the conformal mode propagating in the RG flow.\footnote{By renormalising $\dot{\Delta}$ this corresponds to dropping the cross-contracted piece.}
Secondly, a value of $j=-1/D$ allows only traceless fluctuations to propagate in the RG flow.

To see why $j\to\infty$ only carries the conformal mode in the RG flow, let us rewrite the metric to bring a scale factor, $e^{\sigma}$, outside of a fixed-scale metric, $\tilde{g}_{\mu\nu}$:
\begin{equation}\label{conformalmetric}
 g_{\mu\nu} = \tilde{g}_{\mu\nu}e^{\sigma}.
\end{equation}
We now see that
\begin{equation}
\label{d-sigma}
 \frac{\delta S}{\delta \sigma} = g_{\mu\nu}\frac{\delta S}{\delta g_{\mu\nu}}.
\end{equation}
This tells us that, if we only use the two-traces structure, then only the conformal mode propagates in the flow equation. Therefore this limit for the flow equation is the so-called conformal truncation, or conformally reduced gravity model \cite{Machado:2009ph,Reuter:2008qx,Reuter:2008wj,Bonanno:2012dg,Dietz:2015owa}.

Conversely, since any symmetric two-tensor can be split uniquely into its trace and trace-free part:
\be 
T^{\rho\sigma} = g^{\rho\sigma} T/D + T^{\rho\sigma}_{\rm trace-free}\,,
\ee
and since 
\be 
\left(g_{\mu(\rho}g_{\sigma)\nu}+jg_{\mu\nu}g_{\sigma\rho}\right)g^{\rho\sigma} = g_{\mu\nu} (1+jD)\,,
\ee
the pure trace part of any variation is excluded from the flow for $j=-1/D$.
%
%
This choice therefore decouples the cosmological constant from the flow equation at the classical level, leaving it as a pure integration constant that does not mix with other scales. Therefore this limit for the flow equation is related unimodular gravity \cite{Einstein1919,Unruh1989a,Eichhorn:2015bna,Saltas:2014cta}.
We will not discuss these special cases further.

Finally, it will be helpful to note that the flow equation at the classical level, \eqref{flowratio}, 
can be written compactly as \cite{Morris:1998kz}
\be 
\label{a0}
\dot{S} = -a_0[S,\Sigma]\,,
\ee
where $a_0$ is symmetric bilinear. Writing $S= \int_x \! \sqrt{g}\, \cL$, where the Lagrangian density $\cL$ is a scalar (and likewise relate $\hat{S}$ to $\hat{\cL}$), we can alternatively write this as a symmetric bilinear map between Lagrangians:
\be 
\label{ca0}
\dot{\cL} = -\ca_0[\cL,\cL-2\hat{\cL}]\,.
\ee

\subsection{Dimensional analysis}
\label{sec:dimensions}

Further insight into the gravity flow equation can be gained from dimensional analysis (using the usual so-called engineering dimensions). First consider the scalar case. The (mass) dimension of a scalar field is $(D-2)/2$, from \eqref{scalarSeed} for example. Since the action must be dimensionless,  $\dot{\Delta}$ expressed as a differential operator (or in momentum space) then has dimension $-2$, from \eqref{scalarSdot} for example, consistent with regarding $\Delta$ as an effective propagator.
 
Next, consider the gauge theory case.
Expressing the covariant derivative as $D_\mu = \partial_\mu - i g A_\mu$, in the way appropriate for perturbative quantum field theory with canonically normalised kinetic term, the dimensional assignments in $D$ space-time dimensions are the same. However expressing the covariant derivative as \eqref{gaugeCovariantDerivative} already leads to a difference outside $D=4$ dimensions, which as we will see is instructive to understand. Now the gauge field must always have dimension $1$ and thus to keep the action dimensionless we recover from \eqref{rescaleact} that $[g^2]=4-D$. If we tried to use the definition $\Sigma = S - 2\hat{S}$ that we used in scalar field theory, we would have to have $[\Delta] = 2-D$ to balance dimensions in  \eqref{gaugeflow}. This is actually consistent with regarding $\Delta$ as an effective propagator since indeed $g^2 c/p^2$ has this dimension, the factor of $g^2$ coming from the non-canonical normalisation of the kinetic term in \eqref{rescaleact}. Once $g$ runs with $\Lambda$ however the flow equation will no longer be consistent because $\dot{\Delta}$ then has a $1/p^2$ pole so is no longer quasi-local. This is a problem because the non-locality will in turn be inherited by solutions $S$ for the Wilsonian action.
The change in definition of $\Sigma$ to $\Sigma_g$, as in \eqref{Sigma-g}, not only ensures a sensible gauge invariant perturbative expansion but also makes $[\Sigma_g]=4-D$ (and $[\hat{S}]=4-D$ consistent with the fact that it does not contain $g$), and thus from \eqref{gaugeflow} allows $[\dot{\Delta}] = -2$ consistent with it playing the r\^ole of a canonically normalised effective propagator, and ensuring that the flow equation remains quasi-local.

Finally let us return to the gravity flow equation. Since $[g_{\mu\nu}]=0$, and the actions are dimensionless, the putative ``effective propagator'' in \eqref{c.c.} and \eqref{t.t.} has dimension 
\be 
\label{dimDelta}
[\Delta]=-D\,.
\ee 
Again, this is to be expected. In $D=4$ dimensions we see that the effective propagator will have to take the form 
\be 
\label{prop-Weyl}
\Delta(p^2) = \frac{c(p^2/\Lambda^2)}{p^4}\,,
\ee
for some function $c$,
at a fixed point, recovering the fact that classically this will involve a four-derivative $R^2$-type action around the Gaussian fixed point (\aka free gravitons). Perturbative quantum gravity based on such an action can be renormalisable and asymptotically free but suffers from problems with unitarity \cite{Stelle:1976gc,Adler:1982ri,Codello:2008}. In order to implement universality as widely as possible we want to avoid having to restrict the form of the cutoff profile function $c(p^2/\Lambda^2)$ beyond normalisation $c(0)=1$, smoothness (that is being infinitely differentiable) and the requirements that will eventually be placed on its asymptotic behaviour to ensure UV finiteness of the flow equation at the quantum level. In this case, for $\dot{\Delta}$ to remain quasi-local however, we will need to restrict the cutoff profile to satisfy $c'(0)=0$. We will pursue this solution for the flow equation in sec. \ref{sec:bifixed}. Since it can be arranged that there is one asymptotically free coupling $\lambda_W$ which is proportional to the inverse coefficient of the square of the Weyl curvature  (and another coupling $\omega\to\omega_* \approx -0.0228$ in the $\Lambda\to\infty$ limit) \cite{Avramidi1985,deBerredoPeixoto:2004if,Codello2006,Codello:2008} we will refer to this approach to the flow equation as the ``Weyl scheme''. (The running of these couplings follow from logarithmic UV divergences and thus can be expected to be universal, independent of regularisation and renormalisation scheme.)

If the Lagrangian contains the Einstein-Hilbert term $-R/(16\pi G)$ then Newton's constant has dimension 
$[G]=2-D$. If we want the effective propagator to be derived from this term, it will now be $\Delta\sim G c/p^2$,  and indeed again has dimension $-D$. Once $G$ runs with $\Lambda$ however, such a term is once more unacceptable. 
Again this problem is avoided by redefining $\Sigma$, this time to $\Sigma = 4S/M^2 - 2\hat{S}$ where $M$ is the reduced Planck mass: $M^2 = 1/(8\pi G)$, and allowing 
\be 
\label{prop-Einstein}
\Delta(p^2) = \frac{c(p^2/\Lambda^2)}{p^2}\,,
\ee
 corresponding to a canonically normalised kinetic term. The classical limit corresponds to $M\to\infty$ such that we retain only $S=M^2 S_0/4$ in the expansion \eqref{mass-expansion}, and again we then relabel $S_0$ as $S$. Again this corresponds to building the theory around the Gaussian fixed point (for canonically normalised kinetic term, in the limit $M\to\infty$, it again describes free gravitons), however this time with the irrelevant perturbation, parametrized by $G=1/(8\pi M^2)$, built in. Note that the actions however then have mass dimension $[S_0]=[\hat{S}]=-2$. We will refer to this form of flow equation as the ``Einstein scheme'' and give more detail on this in sec. \ref{sec:biEHfixed}.


Either way at the classical level the flow equation will reduce to \eqref{a0}, \ie \eqref{flowratio}, where the individual terms are defined in \eqref{c.c.} and \eqref{t.t.}. 
From here on, we will exclusively consider space-time dimension $D=4$. 
Since the classical action can be dimensionless or dimension -2 depending on whether we use the Weyl or Einstein scheme, 
the dimension of the Lagrangian is respectively $[\cL]=[\hat{\cL}]=\ell=4$~or~2.

\section{Background-independent expansion of the effective action}\label{sec:bieffact}

In the background-independent computation, $\cL$ (and likewise $\hat{\cL}$) can be organised by expanding in a basis of local diffeomorphism invariant scalar operators $\cO_{d}$ of increasing even engineering mass dimension  $d=2i$:
\be 
\label{lagrangian-expansion}
\cL = \sum_{i=0}^\infty \sum_{\alpha_i} g_{2i,\alpha_i}\, \cO_{2i,\alpha_i}\,,
\ee
where the operators contain only the metric and space-time derivatives,\footnote{{\it N.B.} we use position space, since a momentum space only makes sense in a translation invariant background.} and the $\alpha_i$ are extra labels which we usually suppress, but which are needed when there is more than operator of the given dimension. The couplings $g_d$ are therefore of dimension $\ell-d$. 

Note that since the metric has dimension zero, the operator dimension just counts the number of space-time derivatives required to construct it.
 Thus the lowest dimension operator is just the unit-operator, $\cO_0=1$, 
 whose associated coupling $g_0(\Lambda)$ we can loosely regard as associated to the effective cosmological constant. (In general this coupling runs with $\Lambda$. It therefore does not correspond to the cosmological constant $\lambda_C$ until the functional integral is completed by sending $\Lambda\to0$. Furthermore since the coefficient of $\sqrt{g}$ is actually $\lambda_C/(8\pi G)$, in the Weyl scheme we must still also compute the effective Planck mass, then finally $\lambda_C=g_0(0)/M^2$.) The next higher dimension operator is $\cO_2=-2R$. We include the minus sign gained through Wick rotation from Minkowski signature and the factor two for canonical normalisation of the graviton kinetic term. In the Weyl scheme its coupling $g_2$ will provide the effective Newton coupling or Planck mass in the limit $\Lambda\to0$, through $g_2 = 1/(32\pi G) = M^2/4$. In the Einstein scheme we already have a (running) reduced Planck mass $M$ but which we so far have not defined precisely. To do this a natural refinement of the scheme is to define $M^2(\Lambda)$ to be the coefficient of $-R/2$ at cutoff-scale $\Lambda$.\footnote{For further discussion of schemes in Wilsonian, and also holographic contexts, see ref. \cite{Lizana:2015hqb}.}
Thus in the Einstein scheme, recalling that we have defined \new{the classical part of the action by }$S= M^2 S_0/4$, defining $M$ in this way, we impose that $g_2=1$. At dimension 4, there are two linearly independent operators which may be taken to be $\cO_{4,1}=R^2$ and $\cO_{4,2}=R^{\mu\nu}R_{\mu\nu}$.\footnote{Since we tacitly assume a space with no boundary throughout the paper, the third possibility $R_{\mu\nu\rho\sigma}R^{\mu\nu\rho\sigma}$ is  linearly related to the other two up to the generalized Gauss-Bonnet topological invariant (in $D=4$ dimensions) which thus decouples from the other terms in the flow equation. We will not consider it further in this paper.} At dimension 6, for the first time we have operators with explicit covariant derivatives appearing (for example $R\nabla^2 R$), and also for the first time we have operators containing more than two curvature factors that thus do not contribute to the two-point vertex (for example $R^3$).

Given the quasi-local form of the flow equation, whatever quasi-local form we choose for $\hat{S}$, we can solve for the general form of the classical action iteratively, starting from the lowest dimension operators. 

Let us illustrate this with the specific forms \eqref{c.c.} and \eqref{t.t.}.\footnote{However, we keep the discussion at a general level. In secs. \ref{sec:bifixed} and \ref{sec:biEHfixed} we will give concrete examples.}  In this case the requirement of quasi-locality enforces that the kernel inserts a Taylor series in $\nabla^2$: 
\be 
\label{kernel-taylor}
\dot{\Delta}(-\nabla^2) = \sum^\infty_{k=0} \frac{1}{k!}\, \dot{\Delta}^{(k)}(0)\left(-{\nabla^2}\right)^k\,,
\ee
(the coefficients $\dot{\Delta}^{(k)}(0)$ depend on the scheme and are examined in more detail in  secs. \ref{sec:bifixed} and \ref{sec:biEHfixed}). We first note that $\ca_0[\cO_d,\cO_{d'}]$ can also be expanded in operators of definite dimension. Indeed, given that 
\[
\frac{\delta}{\delta g_{\mu\nu}}\int_x \! \sqrt{g}\, \cO_{d}
\]
is also dimension $d$, we see that $\ca_0$ in the flow equation \eqref{ca0} has the property that 
\be 
\label{ca0k}
\ca_0[\cO_{d_1},\cO_{d_2}] = \sum^\infty_{k=0} 
 \ca^k_0[\cO_{d_1},\cO_{d_2}]\,,
\ee 
where $\ca^k_0[\cO_{d_1},\cO_{d_2}]$ is a linear combination of operators $\cO_{d}$ with dimension $d=d_1+d_2+2k$, and is proportional to $\dot{\Delta}^{(k)}(0)$. Since $d,k\ge 0$, a coupling $g_d$ can only appear in the flow of couplings $g_{d'}$ where $d'\ge d$. Therefore, as claimed, we can solve iteratively for all the couplings ordered according to the dimension of the associated operator.

In particular, the effective cosmological constant $g_0$ obeys a closed equation:
\be 
\label{g0-flow}
\dot{g}_0=g_0(2\hat{g}_0-g_0)\, \ca^0_0[1,1]\,,
\ee
which is readily solved. ($\ca^0_0[1,1]\propto \dot{\Delta}(0)$ is just a number times a power of $\Lambda$.) 
 Plugging $g_0(\Lambda)$ into the flow of $g_2$:
\be 
\label{g2-flow}
\dot{g}_2 = 2 (g_0 \hat{g}_2 +\hat{g}_0 g_2 -g_0 g_2)\, \frac{\ca^0_0[\cO_2,1]}{\cO_2} \,,
\ee
allows this to be solved, yielding $g_2(\Lambda)$. (Note that the final term 
again is proportional to $\dot{\Delta}(0)$ and is a number times a power of $\Lambda$.) Note the dimension-two term $\propto\ca^2_0[1,1]$ which would have been {\it a priori} expected, vanishes. In fact 
\be 
\label{k-1}
\ca^k_0[\cO_d,1] =0 \qquad\forall k>0\,,
\ee
since $\nabla_\mu g_{\alpha\beta}=0$. Armed with $g_0$ and $g_2$, the two couplings $g_{4,1}$ and $g_{4,2}$ can now be solved for \etc

\new{As we will see now, t}he seed action couplings $\hat{g}_d$ are subject to some constraints, which turn out to be sufficient to determine the $\hat{g}_{d,\alpha_d}$ (up to a binary decision in the Weyl scheme) for all $d\le4$.

As remarked at the end of sec. \ref{sec:scalars}, we want to be able to construct a fixed point action $S$ and then flow out of this to form the continuum limit (or in the effective field theory context flow into this to form an approximate description valid at energies less than the Planck mass). This is only possible if the seed action contains no scale apart from $\Lambda$. Therefore by dimensions $\hat{g}_d \propto \Lambda^{\ell-d}$ where the coefficients are pure numbers. 

For convenience we impose that when expanded around a flat background, the $\hat{S}$ and $S$ two-point vertices are equal at the fixed point. Thus the fixed point values of the $g_{d,\alpha_d}$ are subject to constraints. For $d\le 4$, this is simply that the fixed point values $g_{d,\alpha_d} = \hat{g}_{d,\alpha_d}$. (For $d>4$, this is only true for the operators containing only two curvature factors.) From the flow equation this imposes further constraints on the $\hat{g}_{d,\alpha_d}$.

Before turning to the computations in the two different schemes, it is helpful to note that
\be 
\label{0-1}
\ca_0[\cO_d,1]=\ca^0_0[\cO_d,1] = \frac{1}{8}(d-4)(1+4j)\dot{\Delta}(0)\,\cO_d\,.
\ee
To see this, we note that from \eqref{a0}, \eqref{flowratio}, \eqref{c.c.} and \eqref{t.t.}, we have
\be
a_0\left[S,\int_x\!\! \sqrt{g}\,\right] = -\frac{1}{4} (1+4j)\dot{\Delta}(0)\int_x\!\! g_{\mu\nu} \frac{\delta S}{\delta g_{\mu\nu}}\,.
\ee
But from \eqref{d-sigma} we know that the last \new{factor} just counts powers of $g_{\mu\nu}$. Equations \eqref{k-1} and \eqref{0-1} provide explicit values for all the bilinears involving $\cO_0$.

\subsection{Effective action in the Weyl scheme}\label{sec:bifixed}

We start by solving the constraints on the seed action couplings $\hat{g}_d$, and thus through the flow equation also compute the fixed point action. Since the fixed point values $g_d=\hat{g}_d$ for $d\le 4$, and since\new{,} in the Weyl scheme\new{,} we have $\dot{g}_d =(4-d) g_d$, \eqref{g0-flow} and \eqref{g2-flow}  already determine $\hat{g}_0$ and $\hat{g}_2$. From \eqref{g0-flow} and \eqref{0-1} we find $\hat{g}_0=0$ or $\hat{g}_0=-8/(1+4j) \dot{\Delta}(0)$. Both these solutions in \eqref{g2-flow} imply that $\hat{g}_2 =0$. 

The couplings $\hat{g}_{4,\alpha}$ are pure numbers that at first sight are undetermined. From \eqref{ca0} and \eqref{k-1},
the $g_{4,\alpha}$ satisfy at the fixed point:
\be 
\label{4-flow}
\dot{g}_{4,1} R^2 + \dot{g}_{4,2} R^{\mu\nu}R_{\mu\nu} = 4g^2_2 \ca^0_0[R,R]  +2g_0g_{4,1}\ca^0_0[R^2,1] +2g_0g_{4,2}\ca^0_0[R^2_{\mu\nu},1]\,.
\ee
Since $\dot{g}_{4,\alpha}=0$ the left hand side vanishes.
\new{One} would \new{usually expect this to }force constraints\new{,} however\new{,} remarkably\new{,} the right hand side vanishes already for any ${g}_{4,\alpha}$,  as follows from \eqref{k-1} and $g_2=0$. Thus so far the fixed point couplings $g_{4,\alpha}=\hat{g}_{4,\alpha}$ can be any pure number.

(If the right hand side had not vanished, for example if $g_2\ne0$, we would have found $\dot{g}_{4,\alpha} = -2r$, where $r$ is a non-vanishing pure number. This has been disallowed by the fixed point condition and quasi-locality. Indeed $\dot{g}_{4,\alpha} = -2r$
would imply $g_{4,\alpha} = r \ln(\mu^2/\Lambda^2)$. However at a fixed point $\mu$ cannot be a separate scale. Neither can $\mu$ inherit a scale from modifying the operators themselves, for example by replacing $R^2$ by  $R^2\ln(R/\Lambda^2)$ or $R\ln(-\nabla^2/\Lambda^2)R$, without violating quasi-locality. )

In fact, in order to fully develop the Weyl scheme, we would have to isolate the asymptotically free coupling $\lambda_W$ and perform an expansion as in \eqref{rescaleactseries} (again to avoid the problems with quasi-locality that would follow from $\Delta\propto \lambda_W$ once $\lambda_W$ runs with $\Lambda$). Then in order to define $\lambda_W$ through the renormalisation scheme we would have to fix the numerical value of $g_{4,2}$. The normalisation implied by the definition of $\lambda_W$ used in refs. \cite{Codello2006,Codello:2008} results in $g_{4,2}=1$, however in order to canonically normalise the effective propagator and kinetic term of the graviton (in sec. \ref{sec:fbfixed}) we choose a different normalisation and set instead $g_{4,2}=2$. Following 
the fixed point analysis in refs. \cite{Codello2006,Codello:2008} the ratio of the $\hat{g}_{4,\alpha}$ is determined by $\omega_*$. In this way both the $\hat{g}_{4,\alpha}$ are in fact already determined. 

Thus the couplings $\hat{g}_d$ for $d\le4$ are all determined up to a binary decision for $\hat{g}_0$. Choosing the simplest possibility $\hat{g}_0=0$, we have thus shown that the seed action is given by
\be \label{startaction}
\hat{S} = 2\!\int_{x} \sqrt{g}\left(R_{\mu\nu}R^{\mu\nu}+sR^2 +\cdots\right)\,,
\ee
where $s$ is a number determined by $\omega_*$ (in fact $s=-(1+\omega_*)/3$ \cite{Codello2006,Codello:2008}) and the ellipsis stands for operators of higher dimension with their associated couplings; those with only two curvature factors will be needed in order to ensure equality of the two-point vertex with the (classical) fixed point $S$ when expanded around a flat background. 

In fact as we will see in sec. \ref{sec:fbfixed}, this determines the seed Lagrangian to be
\be 
\label{Shat-4}
\hat{\cL} = 2R_{\alpha\beta}\,c^{-1}\!(-\nabla^2/\Lambda^2)\, R^{\alpha\beta} + 2sR\, c^{-1}\!(-\nabla^2/\Lambda^2)\, R\,,
\ee
where $c^{-1}$ is the inverted ultraviolet cutoff function. We have the option (by universality) to include more operators providing they contain at least three curvature factors, however we stick with this simplest possibility. The classical fixed point Lagrangian $\cL$ takes the same form as \eqref{Shat-4} for the quadratic curvature terms, but is complemented by an infinite series of further operators which include at least three curvature factors.

Plugging \eqref{startaction} for the fixed point $S$ and $\hat{S}$ back into the flow equation \eqref{a0}, equivalently \eqref{ca0}, we derive
\be 
\label{R4-flow}
\dot{\mathcal{L}} = 4\,\ca_0[R_{\mu\nu}R^{\mu\nu},R_{\alpha\beta}R^{\alpha\beta}] +8s\,\ca_0[R_{\mu\nu}R^{\mu\nu},R^2]+4s^2\ca_0[R^2,R^2]+\cdots\,,
\ee 
where now the ellipsis stands for terms where $\ca_0$ contains at least one operator of dimension $d>4$. Thus we see from \eqref{ca0k}, that \eqref{startaction} will induce operators $\cO_d$ of dimension $d=8,10,12,\cdots$ and in fact uniquely determine the fixed point couplings $g_{8,\alpha_8}$ and $g_{10,\alpha_{10}}$. (The couplings associated to higher dimension operators, starting with $g_{12,\alpha_{12}}$, receive contributions from these $d=4$ operators but also contributions from $d\ge\new{8}$ operators.)


%
%
%
%
To calculate these fixed point couplings we use the functional derivatives of the action terms:
\begin{eqnarray}\label{derivRmnRmn}
 \frac{\delta}{\delta g_{\mu\nu}}\int_{x}\sqrt{g}R_{\alpha\beta}R^{\alpha\beta} & = & \sqrt{g}\left(\frac{1}{2}g^{\mu\nu}R_{\alpha\beta}R^{\alpha\beta}-2R^{\mu}_{\ \alpha}R^{\nu\alpha} \right.\nonumber \\ & &\left. -\nabla^2 R^{\mu\nu}-\frac{1}{2}g^{\mu\nu}\nabla^2 R+\new{2\nabla_\alpha\nabla^{(\mu} R^{\nu)\alpha}}\right),
\end{eqnarray}
\begin{equation}\label{derivR2}
 \frac{\delta}{\delta g_{\mu\nu}}\int_{x}\sqrt{g}R^2 = \sqrt{g}\left(\frac{1}{2}g^{\mu\nu}R^2 - 2RR^{\mu\nu}+2\nabla^{\mu}\nabla^{\nu}R - 2g^{\mu\nu}\nabla^2 R\right)
\end{equation}
(where we have used the Bianchi identity $\nabla_{\mu}R^{\mu\nu}=\half \nabla^\nu R$).
Thus we find
\begin{eqnarray}
\label{RR-RR}
2\ca_0[R_{\mu\nu}R^{\mu\nu},R_{\alpha\beta}R^{\alpha\beta}] & = & 
R_{\alpha\beta}R^{\alpha\beta}\KoL R_{\gamma\delta}R^{\gamma\delta}
-4R_{\alpha\beta}R^{\alpha}_{\ \gamma}\KoL R^{\gamma\delta}R^{\beta}_{\ \delta}
-4R_{\alpha\beta}R^{\alpha}_{\ \gamma}\KoL\nabla^2 R^{\beta\gamma} 
\nonumber \\ & & 
+\new{8R_{\alpha\beta}R^{\alpha}_{\ \gamma}\KoL\nabla_{\delta}\nabla^{\beta}R^{\gamma\delta}}
-\nabla^2 R_{\alpha\beta}\KoL\nabla^2 R^{\alpha\beta} -\nabla^2 R\KoL\nabla^2 R 
\\ & &
+\new{4\nabla^2 R_{\alpha\beta}\KoL\nabla_{\gamma}\nabla^{\alpha}R^{\beta\gamma}} - \new{4\nabla_{\alpha}\nabla_{(\beta}R_{\gamma)}^{\ \ \alpha}\KoL\nabla_{\delta}\nabla^{\beta}R^{\gamma\delta}} 
-4j 
\nabla^2 R\KoL\nabla^2 R\,, \nonumber 
\end{eqnarray}
\begin{eqnarray}
\label{R2-R2}
2\ca_0[R^2,R^2]
& = & 
R^2\KoL R^2 - 2R^2\KoL \nabla^2 R - 4RR_{\alpha\beta}\KoL R^{\alpha\beta}R + 8RR^{\alpha\beta}\KoL \nabla_{\alpha}\nabla_{\beta}R 
\nonumber \\ & &
- 4\nabla_{\alpha}\nabla_{\beta}R\KoL\nabla^{\alpha}\nabla^{\beta}R - 8\nabla^2 R\KoL \nabla^{2}R
-36j 
\nabla^2 R \KoL \nabla^2 R\,,
\end{eqnarray}
and
\begin{eqnarray}
\label{RR-R2}
2\ca_0[R_{\mu\nu}R^{\mu\nu},R^2]
& = & 
R^2\KoL R_{\alpha\beta}R^{\alpha\beta}-4RR_{\alpha\beta}\KoL R^{\alpha}_{\ \gamma}R^{\beta\gamma} - 2RR_{\alpha\beta}\KoL \nabla^2 R^{\alpha\beta}
\nonumber \\ & &
+\new{4RR^{\alpha\beta}\KoL\nabla_\gamma\nabla_\alpha R^{\gamma}_{\ \beta}} -\nabla^2 R\KoL R_{\alpha\beta}R^{\alpha\beta}+4\nabla_{\alpha}\nabla_{\beta}R\KoL R^{\alpha\gamma}R^{\beta}_{\ \gamma}\nonumber \\ & &
+2\nabla_{\alpha}\nabla_{\beta}R\KoL\nabla^{2}R^{\alpha\beta}
-\new{4\nabla^\alpha \nabla^\beta R\KoL\nabla_\gamma \nabla_\alpha R^{\gamma}_{\ \beta}} - 3\nabla^2 R\KoL \nabla^2 R
\nonumber \\ & &
-\new{12}j
\nabla^2 R\KoL\nabla^2 R\,. 
\end{eqnarray}
These expressions need to be quasi-local since they are part of the Wilsonian flow \eqref{R4-flow}. If we have  the ``effective propagator'' \eqref{prop-Weyl}
discussed in sec. \ref{sec:dimensions}, then from
\be 
\label{dotD-Weyl}
\dot{\Delta}(p^2) = -\frac{2}{\Lambda^2 p^2} \,c'(p^2/\Lambda^2)\,,
\ee
we see that we require the cutoff profile to be restricted so that $c'(0)=0$ as claimed. Since only $\dot{\Delta}$ depends on $\Lambda$ in the above expressions, integrating up \eqref{R4-flow} amounts to replacing $\dot{\Delta}$ by 
\be 
\int \frac{d\Lambda}{\Lambda} \dot{\Delta}(p^2) = \frac{c(p^2/\Lambda^2)-1}{p^4} = \frac{1}{p^4} \sum^\infty_{k=2} \frac{c^{(k)}(0)}{k!} \pcutarg^k,
\ee
where the integration constant $c(0)=1$ is determined by maintenance of quasi-locality. 
To compute the couplings of the dimension $d=8$ and 10 operators $\cO_d$, we therefore replace $\KoL$ by 
\be 
\label{dDelta-replace}
\int \frac{d\Lambda}{\Lambda} \dot{\Delta} = \frac{1}{2\Lambda^4}\,c''(0) -\frac{1}{6\Lambda^6}\,c'''(0) \nabla^2+O(\nabla^4)\,.
\ee
In order to compare with \eqref{Shat-4}, we combine covariant derivatives in the two-curvature terms in  \eqref{RR-RR} -- \eqref{RR-R2}, recognising that commutators of covariant derivatives yield curvature terms and thus contribute operators containing at least three curvature factors. Thus we deduce that both $\hat{\cL}$ and the fixed point $\cL$ have the following  couplings for their respective $d=8,10$ operators:\footnote{The first line combines contributions from all three \eqref{RR-RR} -- \eqref{RR-R2}. The second line has only one contribution, coming from \eqref{RR-RR}.}
\bea 
\label{8-10-ops}
&&-\left\{ 1+4j+4s(2+3s)(1+3j) \right\} \left[ 
\frac{1}{\Lambda^4}\,c''(0) R \left(-\nabla^2\right)^2 R 
+\frac{1}{3\Lambda^6}\,c'''(0) R \left(-\nabla^2\right)^3 R \right]\nonumber\\
&&\qquad-\frac{1}{\Lambda^4}\,c''(0) R_{\mu\nu} \left(-\nabla^2\right)^2 R^{\mu\nu}
-\frac{1}{3\Lambda^6}\,c'''(0) R_{\mu\nu} \left(-\nabla^2\right)^3 R^{\mu\nu}\,.
\eea
Since 
\be 
c^{-1}\left(-\nabla^2/\Lambda^2\right) = 1 -\frac{1}{2\Lambda^4}\,c''(0) \left(-\nabla^2\right)^2 -\frac{1}{6\Lambda^6}\,c'''(0)\left(-\nabla^2\right)^3+O(\nabla^8)\,,
\ee
(recalling that $c(0)=1$ and $c'(0)=0$), we see that \eqref{8-10-ops} agrees with \eqref{Shat-4} already for the $R^2_{\mu\nu}$ terms, and agrees also for the $R^2$ terms providing 
\be \label{jconstraint}
1+4j+4s(2+3s)(1+3j) = s\,,
\ee
which determines 
\be 
\label{j-Weyl}
j = -\frac{1}{4}\, \frac{1+4s}{1+3s}\,.
\ee
We will see that this constraint indeed arises, in the fixed background computation in sec. \ref{sec:fbfixed}. 
\new{Setting $s=-1/3$, with $j$ a free parameter, would also have solved \eqref{jconstraint}, however we have fixed $s$ to the value set by $\omega_{*}$, as below \eqref{startaction}}.
The remaining $d=8,10$ operators coming from \eqref{RR-RR}--\eqref{RR-R2} after using \eqref{dDelta-replace}, have at least three factors of curvature and thus appear in the fixed point $S$ but not in $\hat{S}$.

%
%
%

\subsubsection{Flowing away from the fixed point with dimensionful couplings}\label{sec:biaway}

Any operator added to $S$ with a coupling containing a dimensionful parameter other than $\Lambda${,} will perturb the theory away from the fixed point. At the classical level two such operators are distinguished, namely $\cO_0 =1$ and $\cO_2 = -2R$, since they are relevant eigenoperators and thus generate flow away from the fixed point. We already have the corresponding flow equations in \eqref{g0-flow} and \eqref{g2-flow}. Using the fact that the corresponding fixed point and seed-action couplings vanish,   we have  for the general flows
\be
\label{relevant-dirs}
\dot{g}_0 = \alpha\,{g_0^2}/{\Lambda^4}\,,\qquad 
\dot{g}_2 = \alpha\,{g_0g_2}/{\Lambda^4}
\ee
where, using \eqref{0-1}, \eqref{j-Weyl} and \eqref{dotD-Weyl}, we compute the dimensionless parameter $\alpha = s\,c''(0)/(1+3s)$.  At the linearised level, the couplings $g_0$ and $g_2$ do not flow. Since they have dimension $4$ and $2$ respectively the dimensionless couplings $\tilde{g}_0 =g_0/\Lambda^4$ and $\tilde{g}_2 = g_2/\Lambda^2$ therefore do indeed correspond to relevant directions shooting out from the fixed point. In the limit $\Lambda\to0$, and at the classical level in which we are working, $g_0$ and $g_2$ should provide the physical cosmological constant, and physical Newton constant or Planck mass, as already discussed in sec. \ref{sec:bieffact}. 

In general\new{,} the fact that the fixed point and seed-action couplings coincide for $d\le4$ means that the flow equation for perturbations in these couplings contains no linear terms (or equivalently cross-terms between these and the fixed point values). To see this, let  $\Delta \cL$ contain such  perturbations away from the fixed point solution, then from \eqref{ca0}, 
\bea
\sum_{i=0}^2 \sum_{\alpha_i} \dot{g}_{2i,\alpha_i}\, \cO_{2i,\alpha_i} &\in& -\ca_0[\cL+\Delta\cL,\cL+\Delta\cL-2\hat{\cL}] \nonumber\\
&\in& \ca_0[\cL,\cL]-\ca_0[\Delta\cL,\Delta\cL]\,.
\eea
Since at the fixed point $\dot{g}_{4,\alpha}=0$, using \eqref{4-flow} we thus read off the flow for the $d=4$ couplings away from the fixed point: 
\be 
\label{D4-flow}
\dot{g}_{4,1} R^2 + \dot{g}_{4,2} R^{\mu\nu}R_{\mu\nu} = -4g^2_2 \ca^0_0[R,R] = 2\dot{\Delta}(0)\, g_2^2\left(R_{\mu\nu}R^{\mu\nu}+jR^2\right)\,,
\ee
while from \eqref{dotD-Weyl} we see that $\dot{\Delta}(0)=-{2\,c''(0)}/{\Lambda^4}$.

Note that it is straightforward to  solve the  $g_0$ flow equation in \eqref{relevant-dirs}. Substituting the result into the flow for $g_2$ allows $g_2$ to be straightforwardly solved for. Substituting $g_2$ into the above equation then allows us to solve straightforwardly for $g_{4,1}$ and $g_{4,2}$. Continuing in this way we can iteratively construct and solve the flows for operators $\cO_d$ up to any desired dimension $d$. 

Note that if $g_2\ne0$ then \eqref{D4-flow} implies in particular that the coupling $g_{4,2}$ now runs even at the classical level. In fact as we discussed above \eqref{startaction}, in a full development of the Weyl scheme we would set $g_{4,2}=2$, and expand in a power series in the coupling $\lambda_W$. The running would then be accounted for in a classical contribution to the running of $\lambda_W$. However inclusion of an Einstein-Hilbert term adds an $O(p^2)$ term to the graviton propagator and therefore it would be more natural to generalise the flow equation to incorporate an ``effective propagator'' of form $\sim 1/(p^4 + aM^2p^2)$ (where $a$ is some dimensionless coefficient). We leave this line of investigation for future research.

\subsection{Effective action in the Einstein scheme}\label{sec:biEHfixed}

As already sketched in sec. \ref{sec:dimensions}, in order to build a flow equation adapted to the Einstein-Hilbert action, we need to scale out Newton's constant $G$ so that it does not appear in the ``effective propagator'' $\Delta$. The action then has the following weak coupling expansion, similar to (\ref{rescaleactseries}):
\begin{equation}
\label{mass-expansion}
 S = \frac{1}{\tilde\kappa} S_{0} + S_{1} + \tilde\kappa S_{2} + \tilde\kappa^2 S_{3}+\cdots
\end{equation}
where $\tilde\kappa=32\pi G$ also counts powers of $\hbar$, and since $\tilde{\kappa}= 4/M^2$ it also an expansion in $1/M^2$, where $M$ is the reduced Planck mass. Thus again $S_i$ is the contribution at the $i^{\rm th}$ loop level, with $S_0$ being purely classical. The coefficient of $\cO_2 = -2R$  in $S$ is set at $g_2=1$ thus defining precisely what we mean by $G(\Lambda)$, equivalently $M(\Lambda)$, but with the consequence that these  run with $\Lambda$ (in general and certainly at the quantum level). Therefore the physical values are only assured in the limit $\Lambda\to0$. Notice that the actions $S_n$ thus have mass dimension $2n-2$, and corresponding Lagrangian densities $\cL_n$ have mass dimension $2n+2$. The flow equation is still \eqref{a0} but now we set $\Sigma = \tilde{\kappa} S-2\hat{S}$. Therefore $\Sigma$ and $\hat{S}$ have mass dimension -2. Let us briefly also consider the quantum part of the flow equation \eqref{full-flow}; it is a linear operator $a_1$ acting on $\Sigma$, thus the full flow equation can be written compactly as \cite{Morris:1998kz}
\be 
\dot{S} = -a_0[S,\Sigma] +a_1[\Sigma]\,.
\ee
Substituting \eqref{mass-expansion} we see that
\bea 
&&\frac{1}{\tilde\kappa} \dot{S}_{0} + \dot{S}_{1} + \tilde\kappa \dot{S}_{2} 
 + \tilde\kappa^2 \dot{S}_{3}+\cdots +\beta \left(
-\frac{1}{\tilde{\kappa}^2} S_0 + S_2 +2\tilde{\kappa} S_3 +\cdots \right) = -\frac{1}{\tilde\kappa}a_0[S_0,S_0-2\hat{S}]
\nonumber\\ 
&&  -2a_0[S_0-\hat{S},S_1]+a_1[S_0-2\hat{S}]
 +\tilde{\kappa}\left(-2a_0[S_0-\hat{S},S_2]-a_0[S_1,S_1]+a_1[S_1]\right)+\cdots\,.
\eea
The classical equation is recovered in the limit $\tilde{\kappa}\to0$, equivalently the $M\to\infty$ limit, where we do not expect it to run. Therefore we find for the classical flow
\be 
\label{classical-Einstein-flow}
\dot{S}_{0} = -a_0[S_0,S_0-2\hat{S}]\,.
\ee
The quantum corrections at the $n^{\rm th}$ loop level can be consistently separated by equating coefficients of $\tilde{\kappa}^{n-1}$. At the same time we see that the beta function must therefore take the general form
\begin{equation}
\label{beta}
 \beta := \Lambda\partial_{\Lambda}\tilde\kappa = \beta_{1}\Lambda^2\tilde\kappa^2 + \beta_{2}\Lambda^4\tilde\kappa^3 + \cdots\,.
\end{equation}
The powers of $\Lambda$ are included by dimensions so that the $\beta_i$ are dimensionless. In the case that $\tilde{\kappa}$ is the only independent dimensionful parameter, the $\beta_i$ will be pure numbers. The formula \eqref{beta} concurs with perturbative expectations (as can be confirmed by expanding $-\sqrt{g} R/(16\pi G)$ about a flat background as in
\eqref{h}, normalising the kinetic term by writing $h_{\mu\nu} = \tilde{h}_{\mu\nu} \sqrt{\tilde{\kappa}}$, and drawing Feynman diagrams). Writing in dimensionless terms by introducing $\kappa= \tilde{\kappa}\Lambda^2=4\Lambda^2/M^2$, the beta function inherits the expected classical term reflecting the fact that it is dimensionally an irrelevant coupling:
\be 
\beta(\kappa) =  \Lambda\partial_{\Lambda}\kappa = 2\kappa +\beta_1 \kappa^2 +\beta_2 \kappa^3 +\cdots\,.
\ee

Now we again consider only the classical limit. Dropping the subscript $0$ on $S$ in \eqref{classical-Einstein-flow} we return to the form of the original flow equation \eqref{a0} as promised, with the only difference being that the actions now have mass dimension -2 (and thus Lagrangian densities have dimension 2).

As discussed in sec. \ref{sec:dimensions}, we can now take the form \eqref{prop-Einstein} for the ``effective propagator'', giving automatically a quasi-local kernel since
\be 
\label{dotD-Einstein}
\dot{\Delta}(p^2) = -\frac{2}{\Lambda^2} \,c'(p^2/\Lambda^2)\,.
\ee
We now deduce the form of the couplings $g_d=\hat{g}_d$ for $d\le4$. Recall that $g_2=1$ is fixed as a normalisation condition.  Thus since we then have $\hat{g}_2=1$, and we maximise universality by avoiding having to impose $\dot{\Delta}(0)=0$, we deduce  from \eqref{g2-flow}, that $\hat{g}_0 = 0$ (and thus at the fixed point $g_0=0$ also).
From \eqref{ca0k}, we see we now have enough information to determine the couplings $g_{4,\alpha}$. Indeed,
\be 
\label{R2-Einstein}
g_{4,1} R^2 + g_{4,2} R^{\mu\nu}R_{\mu\nu} = 4\int \frac{d\Lambda}{\Lambda}\,\ca^0_0[R,R] = -2\frac{c'(0)}{\Lambda^2}\, \left(R_{\mu\nu}R^{\mu\nu}+jR^2\right)\,,
\ee
where we have used the last equality in \eqref{D4-flow}, and noted by \eqref{dotD-Einstein} that now
\be 
\int \frac{d\Lambda}{\Lambda}\, \dot{\Delta}(0) =  \frac{c'(0)}{\Lambda^2}
\ee
(since dimensionful integration constants are not allowed at the fixed point). 

These terms form the beginning of the tower of curvature-squared operators that contribute to the regularised graviton kinetic term when expanded around a fixed background. In sec. \ref{sec:fbEHfixed}, we will see that they continue to appear in the same proportions as in \eqref{R2-Einstein} and thus the seed-Lagrangian takes the form:
\begin{equation}\label{biEHfull}
 \hat{\cL} = -2R + \frac{2}{\Lambda^2}R_{\mu\nu}\,d(-\nabla^2/\Lambda^2) R^{\mu\nu} + \frac{2}{\Lambda^2}jR\, d(-\nabla^2/\Lambda^2) R\,.
\end{equation}
So far, we have shown that $d(0)=-c'(0)$.
As before, we have a choice of whether to include operators containing at least three curvature factors, but take the simplest choice and exclude them. 

In pure gravity, the only relevant perturbation is now $\cO_0=1$, generating a cosmological constant. From \eqref{g0-flow}, \eqref{0-1} and \eqref{dotD-Einstein} we obtain for this flow,
\be 
\label{cosmoFlowE}
\dot{g}_0 = -(1+4j)\frac{c'(0)}{\Lambda^2} g_0^2\,.
\ee
Notice that the flow equation \eqref{g2-flow} is still consistent with the normalisation requirement $g_2=1$, since this together with the seed action couplings ensures that $\dot{g}_2=0$ even with $g_0\ne0$.

\section{Gravity flow equation expanded around fixed background}\label{sec:fbflow}

In the fixed-background approach, we define a metric perturbation as the difference between the metric and a Euclidean background:
\begin{equation}
\label{h}
 h_{\mu\nu}(x) := g_{\mu\nu}(x) - \delta_{\mu\nu}.
\end{equation}
This metric perturbation corresponds to the graviton field.
The inverse metric is then an expansion around a flat background:
\begin{equation}
 g^{\mu\nu}(x) = \delta^{\mu\nu} - h^{\mu\nu}(x) + h^{\mu}_{\ \rho}(x)h^{\nu\rho}(x) + \cdots\,.
\end{equation}
On the right hand side (and from now on) indices are raised and contracted using the background metric $\delta_{\mu\nu}$.
Although we could continue to use position representation, we will find it more useful to Fourier transform into a momentum representation:
\begin{equation} \label{Fourier}
h_{\mu\nu}(x) = \int \dbar p \,{\rm e}^{-ip\cdot x} h_{\mu\nu}(p)\,.
\end{equation}
where we use the shortened notation that
\begin{equation}
\dbar p := \frac{d^{D}p}{(2\pi)^{D}}.
\end{equation}
It is also convenient to define
\begin{equation}
 \delbar(p) := (2\pi)^D \delta(p).
\end{equation}
The action is now defined as an expansion in $n$-point vertices
\begin{eqnarray}\label{ActionSeries}
S & = & \int \dbar p \ \delbar(p)\mathcal{S}^{\mu\nu}(p)h_{\mu\nu}(p)
            + \frac{1}{2}\int \dbar p \ \dbar q \ \delbar(p+q)\mathcal{S}^{\mu\nu\rho\sigma}(p,q)h_{\mu\nu}(p)h_{\rho\sigma}(q) \nonumber \\ & &
            + \frac{1}{3!}\int \dbar p\ \dbar q\ \dbar r \ \delbar(p+q+r)\mathcal{S}^{\mu\nu\rho\sigma\alpha\beta}(p,q,r)h_{\mu\nu}(p)h_{\rho\sigma}(q)h_{\alpha\beta}(r) + \cdots
\end{eqnarray}
We do not include a 0-point function because it has no physical significance.
Since there is only one type of 1-point function, which always has zero for its momentum argument, it is convenient, unless otherwise stated, to write it as
\begin{equation} \label{one-point-S}
\mathcal{S}^{\mu\nu}(0) = \mathcal{S}\delta^{\mu\nu}\,,
\end{equation}
where $\mathcal{S}$ is a constant. The $n$-point functions are obtained by functional differentiation: 
\begin{equation}
\mathcal{S}^{\alpha_1\beta_1\cdots\alpha_n\beta_n}(p_1,\cdots,p_n) = \frac{\delta}{\delta h_{\alpha_1\beta_1}(p_1)}\cdots\frac{\delta}{\delta h_{\alpha_n\beta_n}(p_n)}S\Big|_{h=0}\new{\,.}
\end{equation}
The $n$-point functions are symmetric under exchange of pairs of indices and the associated momentum arguments together,
and under exchange of the indices within a pair. We can \new{re-express} the flow equation (\ref{c.c.}), (\ref{t.t.}) and (\ref{flowratio}), \new{which uses the kernel given in} \eqref{kernel-grav}, \new{in fixed-background form} by noting that
\begin{equation}
\frac{\delta}{\delta g_{\mu\nu}(x)} = \frac{\delta}{\delta h_{\mu\nu}(x)}\,\new{.}
\end{equation}
Fourier transformed into the momentum representation\new{, the flow equation becomes}
\begin{equation}
\dot{S} = \half\int \dbar q\ \dbar r\ \frac{\delta S}{\delta h_{\mu\nu}(-q)}K_{\mu\nu\rho\sigma}(q,r)\frac{\delta \Sigma}{\delta h_{\rho\sigma}(-r)}.
\end{equation}
where $S$, $\Sigma$ and $K$ are all separately momentum conserving.
One then obtains the flow equations at the $n$-point level by functionally differentiating $n$ times, then setting $h_{\alpha\beta}=0$.
Not only the actions, $S$ and $\Sigma$, but also the kernel, $K$, consist of an infinite expansion in metric perturbations, in a spirit similar to (\ref{ActionSeries}).
The $n$-point structure of the kernel differs from that of the actions since there are $n+2$ momentum arguments for each $n$-point function.
Also, the kernel $n$-point expansion begins with $n=0$ rather than $n=1$.  

Expanding the flow equation in powers of $h_{\alpha\beta}$ thus gives a diagrammatic form that looks exactly like that of fig. \ref{fig:Anpoint}, the only difference being that the action terms now generically have one-point vertices, as we noted above. At first sight\new{,} that means that the classical flow of $n$-point functions is no longer closed but rather receives a contribution from a one-point vertex (tadpole) attached to an $(n$+1)-point vertex. This is actually not the case, since such an $(n$+1)-point vertex has a zero momentum argument and can thus be related back to $n$-point vertices via differential Ward identities as we show in the next section. In fact\new{,} a one-point vertex can only arise from a cosmological constant (\ie $\cO_0$) term as we will also demonstrate explicitly in the next section, and we have already seen in the background-independent computation, namely \eqref{0-1}, that attaching such a term just multiplies the other operator $\cO_d$ by a $d$-dependent factor.

Thus the classical $n$-point vertices can be solved for iteratively, \ie once the ($m\!<\!n$)-point vertices have been determined.

\section{Ward identities}

The diffeomorphism invariance of the action allows us to relate the $(n$+1)-point functions of the action to their 
respective $n$-point functions via Ward identities.
Since the kernel is a diffeomorphism covariant bitensor, it is also possible to derive Ward identities for it separately.
The Ward identities for $\dot{S}$ in the flow equation can then be consistently derived either using the usual Ward identity for an action or by the more laborious method of using the Ward identities of $S$, $\Sigma$ and $K$ separately in the flow equation. We have verified explicitly that the results are the same, providing a non-trivial consistency check on our derivations. In sec. \ref{sec:check} we give an example of such a consistency check.

\subsection{Ward identities for an action}\label{sec:WIa}

The variation under diffeomorphisms of the metric perturbation is given by the Lie derivative of the total metric:
\begin{equation} 
\delta h_{\alpha\beta}= \mathsterling_{\xi}\left(\delta+h\right)_{\alpha\beta} = 2(\delta+h)_{\lambda(\alpha}\partial_{\beta)} \xi^\lambda + \xi\cdot\partial h_{\alpha\beta}\,.
\end{equation}
Writing this in momentum space and
requiring the variation in the total action (\ref{ActionSeries}) to be zero gives us the action Ward identities:
\begin{eqnarray}\label{WardIds}
-2p_{1\mu_1} \Sv^{\mu_1\nu_1\cdots\mu_n\nu_n}(p_1,\cdots,p_n) & = &
 \sum_{i=2}^n \pi_{2i} \Big\{\, p_2^{\nu_1}\Sv^{\mu_2\nu_2\cdots\mu_n\nu_n}(p_1+p_2,p_3,\cdots,p_n) \\ &&
+2p_{1\alpha}\delta^{\nu_1(\nu_2}\Sv^{\mu_2)\alpha\mu_3\nu_3\cdots\mu_n\nu_n}(p_1+p_2,p_3,\cdots,p_n)\,\Big\}\,,\nonumber
\end{eqnarray}
where $\pi_{2i}$ is the transposition operator effecting the substitution $p_2,\mu_2,\nu_2\leftrightarrow p_i,\mu_i\nu_i$, and momentum conservation $p_1+\cdots+p_n=0$ is assumed.
The 2-point Ward identity is thus
\begin{equation}\label{Ward1to2}
2p_{\mu}\mathcal{S}^{\mu\nu\rho\sigma}(p,-p) = \mathcal{S}p_{\mu}\delta^{\mu\nu}\delta^{\rho\sigma}-2\mathcal{S}p_{\mu}\delta^{\mu(\rho}\delta^{\sigma)\nu}\,,
\end{equation}
which is only non-zero if the 1-point function, which is momentum-independent, is non-zero.

The 2-point function can thus be split into a transverse momentum-dependent part and a non-transverse momentum-independent part. 
Here we determine the form of the momentum\new{-}independent part by solving the Ward identities, and confirm that they reproduce the cosmological constant part of the action. In secs.  \ref{sec:transverse} and \ref{sec:2pts} we can thus concentrate on transverse two-point functions.

To extract the momentum-independent part of the Ward identity, we first compute the differential Ward identity, for example by putting $n\mapsto n+1$ in \eqref{WardIds} and
choosing the momenta to be $\epsilon,p_1-\epsilon,p_2,\cdots,p_n$. It is easy to see that in the limit $\epsilon\to0$, both sides of \eqref{WardIds} vanish. The $O(\epsilon)$ piece then gives:
\bea 
\label{diffWard}
-2\Sv^{\alpha\beta\mu_1\nu_1\cdots\mu_n\nu_n}(0,p_1,\cdots,p_n) &=& \left(\sum_{i=1}^n p_i^\beta\partial^\alpha_i-\delta^{\alpha\beta}\right) \Sv^{\mu_1\nu_1\cdots\mu_n\nu_n}(p_1,\cdots,p_n)\nonumber\\
&&+2\sum_{i=1}^n\pi_{1i}\,\delta^{\beta(\nu_1}\Sv^{\mu_1)\alpha\mu_2\nu_2\cdots\mu_n\nu_n}(p_1,\cdots,p_n)\,,
\eea
(where in this context, $\partial^\alpha_i$ \new{denotes} differentiation with respect to $p_{i\alpha}$). Thus\new{,} as we claimed at the end of the last section, a vertex with a zero momentum argument is related to vertices with one less leg through the differential Ward identity. 

In fact the tadpole term will involve contraction of $\alpha$ and $\beta$ through the attachment of \eqref{one-point-S}. Then the above equation simply becomes:
\be 
\label{diffWardcontracted}
\Sv^{\ \alpha\mu_1\nu_1\cdots\mu_n\nu_n}_\alpha(0,p_1,\cdots,p_n) = \left(2-n-\half\sum_{i=1}^n p_i\cdot\partial_i\right)\Sv^{\mu_1\nu_1\cdots\mu_n\nu_n}(p_1,\cdots,p_n)\,.
\ee
The differential operator just counts momentum, in the sense that, if we Taylor expand the $n$-point vertex, the differential operator counts the overall power $d$ of momentum in any given term, \ie the dimension of the associated operator $\cO_d$. Thus we recognise that the operator is simply multiplied by a factor involving $(d-4)$ as in \eqref{0-1}. (Matching the $n$ dependence requires also the $m$-point vertices from $\sqrt{g}\cO_0$ and the kernel.)

To extract the momentum-independent part of the Ward identity, we just set all momenta to zero in \eqref{diffWard}:
\begin{equation}\label{Wardmomind}
2\Sv^{\mu_1\nu_1\cdots\mu_n\nu_n}(\underline{0}) = \delta^{\mu_1\nu_1}\Sv^{\mu_2\nu_2\cdots\mu_n\nu_n}(\underline{0})
-2\sum_{i=2}^n\pi_{2i} \,\delta^{\nu_1(\nu_2}\Sv^{\mu_2)\mu_1\mu_3\nu_3\cdots\mu_n\nu_n}(\underline{0})\,.
\end{equation}
These momentum-independent Ward identities allow us to derive the unique form of the zero-momentum part of the action, starting from the 1-point function. Thus we find 
the  2-point function at zero momentum is found to be
\begin{equation}\label{Ward2ptCC}
2\mathcal{S}^{\mu\nu\rho\sigma}(0,0) = \mathcal{S}\delta^{\mu\nu}\delta^{\rho\sigma}-2\mathcal{S}\delta^{\mu(\rho}\delta^{\sigma)\nu}.
\end{equation}
The momentum-independent 3-point function can be written as
\be\label{Ward3ptCC}
2\mathcal{S}^{\mu\nu\rho\sigma\alpha\beta}(0,0,0)  =  2\mathcal{S}\delta^{(\alpha|(\mu}\delta^{\nu)(\rho}\delta^{\sigma)|\beta)}
- \mathcal{S}\delta^{\mu\nu}\delta^{\rho(\alpha}\delta^{\beta)\sigma}
 -2\mathcal{S}^{(\mu|\alpha\rho\sigma}(0,0)\delta^{\beta|\nu)}
 +\mathcal{S}^{\mu\nu\rho\sigma}(0,0)\delta^{\alpha\beta}\,.
\ee
This can then be iterated to any desired $n$-point level.
These structures correspond to the $n$-point structure of $\sqrt{g}$ by itself (see app. \ref{sec:sqrtg}), 
the cosmological constant part of the action.
 
\subsection{Ward identities for a kernel}\label{sec:WIk}

The same principle applies to the kernel, except that the kernel is not diffeomorphism invariant, but rather a covariant bitensor, $K_{\mu\nu\rho\sigma}(x,y)$.
The Lie derivative for the kernel is therefore
\begin{eqnarray}\label{proptrans}
 \mathsterling_{\xi}K_{\mu\nu\rho\sigma}(x,y) & = & \xi(x)\cdot\partial_{x}K_{\mu\nu\rho\sigma}(x,y) + \xi(y)\cdot\partial_{y}K_{\mu\nu\rho\sigma}(x,y) \nonumber \\ && + 2K_{\lambda(\mu|\rho\sigma}(x,y)\partial_{x|\nu)}\xi^{\lambda}(x)+ 2K_{\mu\nu\lambda(\rho|}(x,y)\partial_{y|\sigma)}\xi^{\lambda}(y)\new{\,.}
\end{eqnarray}
The two position coordinates are Fourier transformed separately into momentum space:
\begin{equation}
 K_{\mu\nu\rho\sigma}(x,y) = \int \dbar q\ \dbar r\ e^{-iq\cdot x-ir\cdot y}K_{\mu\nu\rho\sigma}(q,r).
\end{equation}
The kernel is itself an expansion in metric perturbations subject to momentum conservation:
\begin{equation}
 K_{\mu\nu\rho\sigma}(q,r) = \mathcal{K}_{\mu\nu\rho\sigma}(q,r) + \int \dbar p_1\ \delbar(p_1+q+r)\mathcal{K}^{\alpha_1\beta_1}_{\ \ \ \ \mu\nu\rho\sigma}(p_1,q,r)h_{\alpha_1\beta_1}(p_1) +\cdots
\end{equation}
The Ward identities for the kernel then follow in the same way as the Ward identities for the action, except that the right hand side of (\ref{proptrans}) is not zero.
Thus we modify (\ref{WardIds}) to
\begin{eqnarray}\label{kernWard}
&& 2p_{\gamma}'\mathcal{K}^{\gamma\delta\alpha_1\beta_1\cdots\alpha_n\beta_n}_{\ \ \ \ \ \ \ \ \ \ \ \ \ \ \mu\nu\rho\sigma}(p',p_1,\cdots,p_n,q,r) = \nonumber \\
&& -(p'+q)^\delta \mathcal{K}^{\alpha_1\beta_1\cdots\alpha_n\beta_n}_{\ \ \ \ \ \ \ \ \ \ \ \ \mu\nu\rho\sigma}(p_1,\cdots,p_n,q+p',r) \nonumber \\
&& -(p'+r)^\delta \mathcal{K}^{\alpha_1\beta_1\cdots\alpha_n\beta_n}_{\ \ \ \ \ \ \ \ \ \ \ \ \mu\nu\rho\sigma}(p_1,\cdots,p_n,q,r+p') \nonumber \\
&& +2\delta^{\lambda\delta}p'_{(\mu|}\mathcal{K}^{\alpha_1\beta_1\cdots\alpha_n\beta_n}_{\ \ \ \ \ \ \ \ \ \ \ \ \ |\nu)\lambda\rho\sigma}(p_1,\cdots,p_n,q+p',r) \nonumber \\
&& +2\delta^{\lambda\delta}p'_{(\rho|}\mathcal{K}^{\alpha_1\beta_1\cdots\alpha_n\beta_n}_{\ \ \ \ \ \ \ \ \ \ \ \ \ \mu\nu|\sigma)\lambda}(p_1,\cdots,p_n,q,r+p') \nonumber \\
&& -\sum_{i=1}^{n}\pi_{i1}\left\{ p^{\delta}_{1}\mathcal{K}^{\alpha_1\beta_1\cdots\alpha_n\beta_n}_{\ \ \ \ \ \ \ \ \ \ \ \ \mu\nu\rho\sigma}(p'+p_1,p_2,\cdots,p_n,q,r)\right. \nonumber \\
&& \left. +2p'_{\lambda}\delta^{\delta(\alpha_1}\mathcal{K}^{\beta_1)\lambda\alpha_2\beta_2\cdots\alpha_n\beta_n}_{\ \ \ \ \ \ \ \ \ \ \ \ \ \ \ \ \, \mu\nu\rho\sigma}(p'+p_1,p_2,\cdots,p_n,q,r)\right\}.
\end{eqnarray}
The first four terms on the right hand side of (\ref{kernWard}) come from the right hand side of (\ref{proptrans}).
They ensure that all terms in the $\dot{S}$ Ward identities are momentum-conserving by cancelling the momentum-violating contributions originating from the action Ward identities.
In the same way as for the action, we can also extract a differential Ward identity, and Ward identity for the momentum-independent part of the kernel:
\begin{eqnarray}
 \mathcal{K}^{\gamma\delta\alpha_1\beta_1\cdots\alpha_n\beta_n}_{\ \ \ \ \ \ \ \ \ \ \ \ \ \ \ \mu\nu\rho\sigma}(\underline{0}) & = & -\frac{1}{2}\delta^{\gamma\delta}\mathcal{K}^{\alpha_1\beta_1\cdots\alpha_n\beta_n}_{\ \ \ \ \ \ \ \ \ \ \ \ \mu\nu\rho\sigma}(\underline{0}) 
 + \delta^{\lambda(\gamma}\delta^{\delta)}_{\ \ (\mu|}\mathcal{K}^{\alpha_1\beta_1\cdots\alpha_n\beta_n}_{\ \ \ \ \ \ \ \ \ \ \ \ |\nu)\lambda\rho\sigma}(\underline{0})  \\
&& + \delta^{\lambda(\gamma}\delta^{\delta)}_{\ \ (\rho|}\mathcal{K}^{\alpha_1\beta_1\cdots\alpha_n\beta_n}_{\ \ \ \ \ \ \ \ \ \ \ \ \mu\nu|\sigma)\lambda}(\underline{0}) 
-\sum_{i=1}^{n}\pi_{i1}\left\{\delta^{(\gamma|(\alpha_1}\mathcal{K}^{\beta_1)|\delta)\cdots\alpha_n\beta_n}_{\ \ \ \ \ \ \ \ \ \ \ \ \ \mu\nu\rho\sigma}(\underline{0})\right\}\,. \nonumber
\end{eqnarray}
The momentum-independent part of the kernel describes the $n$-point structure for linear combinations of $\frac{g_{\mu(\rho}g_{\sigma)\nu}}{\sqrt{g}}$ and $\frac{g_{\mu\nu}g_{\rho\sigma}}{\sqrt{g}}$, starting with the 0-point function.

\subsection{Consistency of Ward identities}
\label{sec:check}

We can demonstrate the consistency of these Ward identities by applying them in two different ways 
using the fact that the action Ward identities, (\ref{WardIds}), also apply to $\dot{S}$.
Consider the 2-point flow equation for $\dot{S}$:
\begin{equation}\label{Ward2pSdot}
 2p_{\alpha_{1}}\dot{\Sv}^{\alpha_{1}\beta_{1}\alpha_{2}\beta_{2}}(p,-p)=p^{\beta_{1}}\dot{\Sv}^{\alpha_{2}\beta_{2}}(0)-2p_{\lambda}\delta^{\beta_{1}(\alpha_{2}}\dot{\Sv}^{\beta_{2})\lambda}(0).
\end{equation}
We can use the flow equation to expand out $\dot{S}$.
Keeping index and momentum structure explicit for clarity, the 1-point tree-level flow equation can be written as
\begin{equation}\label{1pSdot}
 \dot{\Sv}^{\alpha\beta}(0) = \left(\Sv\left|^{\alpha\beta\mu\nu}(0,0)\mathcal{K}_{\mu\nu\rho\sigma}(0,0)\right|\Sigma\right)^{\rho\sigma}(0) + \Sv^{\mu\nu}(0)\mathcal{K}^{\alpha\beta}_{\ \ \mu\nu\rho\sigma}(0,0,0)\Sigma^{\rho\sigma}(0),
\end{equation}
where the large round brackets indicate anticommutation:
\begin{eqnarray}
&& \left(\Sv\left|^{\alpha\beta\cdots\mu\nu}(p,\cdots,-q)\mathcal{K}_{\mu\nu\rho\sigma}(q,r)\right|\Sigma\right)^{\gamma\delta\cdots\rho\sigma}(p',\cdots,-r)  = \nonumber \\
&& \Sv^{\alpha\beta\cdots\mu\nu}(p, \cdots,-q)\mathcal{K}_{\mu\nu\rho\sigma}(q,r)\Sigma^{\gamma\delta\cdots\rho\sigma}(p',\cdots,-r) + \nonumber \\ 
&& \Sigma^{\alpha\beta\cdots\mu\nu}(p,\cdots,-q)\mathcal{K}_{\mu\nu\rho\sigma}(q,r)\Sv^{\gamma\delta\cdots\rho\sigma}(p',\cdots,-r).
\end{eqnarray}
We can substitute (\ref{1pSdot}) into (\ref{Ward2pSdot}) to get
\begin{eqnarray}\label{Ward2pSdotflow}
 2p_{\alpha_{1}}\dot{\Sv}^{\alpha_{1}\beta_{1}\alpha_{2}\beta_{2}}(p,-p) & = & p^{\beta_{1}}\left(\Sv\left|^{\alpha_2 \beta_2 \mu\nu}(0,0)\mathcal{K}_{\mu\nu\rho\sigma}(0,0)\right|\Sigma\right)^{\rho\sigma}(0)\nonumber \\ 
&& + p^{\beta_1}\Sv^{\mu\nu}(0)\mathcal{K}^{\alpha_2 \beta_2}_{\ \ \ \ \ \mu\nu\rho\sigma}(0,0,0)\Sigma^{\rho\sigma}(0) \nonumber \\ 
&& - 2p_{\lambda}\delta^{\beta_1 (\alpha_2}\left(\Sv\left|^{\beta_2)\lambda}(0,0)\mathcal{K}_{\mu\nu\rho\sigma}(0,0)\right|\Sigma\right)^{\rho\sigma}(0) \nonumber \\ 
&& - 2p_{\lambda}\delta^{\beta_1 (\alpha_2|}\Sv^{\mu\nu}(0)\mathcal{K}^{|\beta_2)\lambda}_{\ \ \ \ \ \mu\nu\rho\sigma}(0,0,0)\Sigma^{\rho\sigma}(0).
\end{eqnarray}
We can test the kernel Ward identity by applying the 2-point flow equation to the left hand side of (\ref{Ward2pSdotflow}) and showing that both sides match after further use of action and kernel Ward identities.
The 2-point tree-level flow equation for $\dot{S}$ is
\begin{eqnarray}\label{general2ptflow}
 \dot{\Sv}^{\alpha_1 \beta_1 \alpha_2 \beta_2}(p,-p) & = & \left(\Sv\left|^{\alpha_1 \beta_1 \mu\nu}(p,-p)\mathcal{K}_{\mu\nu\rho\sigma}(p,-p)\right|\Sigma\right)^{\alpha_2\beta_2\rho\sigma}(p,-p) + \nonumber \\ 
&& \left(\Sv\left|^{\alpha_1\beta_1\mu\nu}(p,-p)\mathcal{K}^{\alpha_2\beta_2}_{\ \ \ \ \mu\nu\rho\sigma}(-p,p,0)\right|\Sigma\right)^{\rho\sigma}(0) + \nonumber \\
&& \left(\Sv\left|^{\alpha_2\beta_2\mu\nu}(p,-p)\mathcal{K}^{\alpha_1\beta_1}_{\ \ \ \ \mu\nu\rho\sigma}(-p,p,0)\right|\Sigma\right)^{\rho\sigma}(0) + \nonumber \\
&& \Sv^{\mu\nu}(0)\mathcal{K}^{\alpha_1\beta_1\alpha_2\beta_2}_{\ \ \ \ \ \ \ \ \ \ \mu\nu\rho\sigma}(p,-p,0,0)\Sigma^{\rho\sigma}(0) + \nonumber \\
&& \left(\Sv\left|^{\alpha_1\beta_1\alpha_2\beta_2\mu\nu}(p,-p,0)\mathcal{K}_{\mu\nu\rho\sigma}(0,0)\right|\Sigma\right)^{\rho\sigma}(0).
\end{eqnarray}

We contract (\ref{general2ptflow}) with $2p_{\alpha_1}$ and apply the action and kernel Ward identities as appropriate.
The first four terms on the right hand side of (\ref{kernWard}) are used to cancel momentum-violating contributions from the action Ward identities.
We will demonstrate the cancellation of momentum-violating terms explicitly in this example.
The first two terms on the right hand side of (\ref{general2ptflow}) only contribute momentum-violating terms from the 2-point Ward identity for an action, which are
\begin{eqnarray}
&& p^{\beta_1}\left(\Sv\left|^{\alpha_2\beta_2\mu\nu}(-p,p)\mathcal{K}_{\mu\nu\rho\sigma}(p,-p)\right|\Sigma\right)^{\rho\sigma}(0)\nonumber \\ 
&& -2p_{\lambda}\delta^{\beta_1 (\mu}\left(\Sv\left|^{\nu)\lambda}(0)\mathcal{K}_{\mu\nu\rho\sigma}(p,-p)\right|\Sigma\right)^{\alpha_2\beta_2\rho\sigma}(p,-p) \nonumber \\
&& +p^{\beta_1}\left(\Sv\left|^{\mu\nu}(0)\mathcal{K}^{\alpha_2\beta_2}_{\ \ \ \ \mu\nu\rho\sigma}(-p,p,0)\right|\Sigma\right)^{\rho\sigma}(0) \nonumber \\ 
&& -2p_{\lambda}\delta^{\beta_1(\mu}\left(\Sv\left|^{\nu)\lambda}(0)\mathcal{K}^{\alpha_2\beta_2}_{\ \ \ \ \mu\nu\rho\sigma}(-p,p,0)\right|\Sigma\right)^{\rho\sigma}(0).
\end{eqnarray}
The next two terms give us contributions from the 1- and 2-point kernel Ward identities. 
After some rearranging of contracted indices, the cancelling contributions are
\begin{eqnarray}
&& -p^{\beta_1}\left(\Sv\left|^{\alpha_2\beta_2\mu\nu}(-p,p)\mathcal{K}_{\mu\nu\rho\sigma}(p,-p)\right|\Sigma\right)^{\rho\sigma}(0) \nonumber \\
&& +2p_{\lambda}\delta^{\beta_1 (\mu}\left(\Sv\left|^{\nu)\lambda}(0)\mathcal{K}_{\mu\nu\rho\sigma}(p,-p)\right|\Sigma\right)^{\alpha_2\beta_2\rho\sigma}(p,-p) \nonumber \\
&& +2p_{\lambda}\delta^{\beta_1(\mu}\left(\Sv\left|^{\nu)\lambda\alpha_2\beta_2}(-p,p)\mathcal{K}_{\mu\nu\rho\sigma}(0,0)\right|\Sigma\right)^{\rho\sigma}(0) \nonumber \\
&& -p^{\beta_1}\left(\Sv\left|^{\mu\nu}(0)\mathcal{K}^{\alpha_2\beta_2}_{\ \ \ \ \mu\nu\rho\sigma}(-p,p,0)\right|\Sigma\right)^{\rho\sigma}(0) \nonumber \\
&& +2p_{\lambda}\delta^{\beta_1(\mu}\left(\Sv\left|^{\nu)\lambda}(0)\mathcal{K}^{\alpha_2\beta_2}_{\ \ \ \ \mu\nu\rho\sigma}(-p,p,0)\right|\Sigma\right)^{\rho\sigma}(0).
\end{eqnarray}
The non-cancelling contributions are
\begin{equation}
 p^{\beta_1}S^{\mu\nu}(0)\mathcal{K}^{\alpha_2\beta_2}_{\ \ \ \ \mu\nu\rho\sigma}(0,0,0)\Sigma^{\rho\sigma}(0) - 2p_{\lambda}\delta^{\beta_1(\alpha_2|}\Sv^{\mu\nu}\mathcal{K}^{|\beta_2)\lambda}_{\ \ \ \ \ \mu\nu\rho\sigma}(0,0,0)\Sigma^{\rho\sigma}(0).
\end{equation}
Since the 1-point kernel Ward identity only gives cancelling terms, the non-cancelling contributions come from the the 2-point kernel Ward identity in this example. 
This just leaves the final term in (\ref{general2ptflow}), which gives us
\begin{eqnarray}
&& p^{\beta_1}\left(\Sv\left|^{\alpha_2\beta_2\mu\nu}(0,0)\mathcal{K}_{\mu\nu\rho\sigma}(0,0)\right|\Sigma\right)^{\rho\sigma}(0)\nonumber \\
&& -2p_{\lambda}\delta^{\beta_1(\alpha_2}\left(\Sv\left|^{\beta_2)\lambda\mu\nu}(0,0)\mathcal{K}_{\mu\nu\rho\sigma}(0,0)\right|\Sigma\right)^{\rho\sigma}(0)\nonumber \\
&& -2p_{\lambda}\delta^{\beta_1(\mu}\left(\Sv\left|^{\nu)\lambda\alpha_2\beta_2}(p,-p)\mathcal{K}_{\mu\nu\rho\sigma}(0,0)\right|\Sigma\right)^{\rho\sigma}(0),
\end{eqnarray}
of which only the last term is a momentum-violating term, coming from the 3-point action Ward identity.
Putting all these terms together, we can match both sides of (\ref{Ward2pSdotflow}). 
Thus we see how momentum-violating contributions from the action Ward identities are cancelled exactly by momentum-violating contributions from the kernel Ward identities, which in turn come from the non-zero Lie derivative of the kernel, as seen in (\ref{proptrans}).

\section{Functional derivatives of the covariantized kernel}\label{sec:kernelpoint}

Like the kernel for gauge theories, the gravity kernel expands out as a series of $n$-point functions. Since we have specified the general form \eqref{kernel-grav}\new{,} we can compute these exactly in terms of the function $\dot{\Delta}$.
It is easy to expand out the momentum-independent part of the kernel as a series in metric perturbations, following app. \ref{sec:sqrtg}. 
For $\dot{\Delta}$, we use the 1-point level as an example.

We start by extracting the $O(h)$ term from $-\nabla^2$ acting on a contravariant tensor, $T^{\rho\sigma}$. 
In momentum representation\new{, our expression at the 1-point level is}\com{I think that the equations are mostly fine as they are, $q$ should not be invoked because these operators only learn about $q$ via the delta function in the kernel (outside of $\KoL$), i.e. derivatives only know what is to the left of them if you integrate by parts, which we are not doing here. Also I don't think that replacing moduli with square brackets is a good idea because we also use square brackets to mean counting mass dimensions elsewhere.}
\begin{equation}\label{dalembertian}
(-\nabla^2)(p,r) T^{\rho\sigma}(-r) = H^{\alpha\beta \ \ \rho\sigma}_{\ \ \gamma\delta}(p,r)T^{\gamma\delta}(-r) h_{\alpha\beta}(p)\,,
\end{equation}
where $H^{\alpha\beta \ \ \rho\sigma}_{\ \ \gamma\delta}(p,r)$ is 
defined by
\begin{multline}\label{dalembert1pt}
H^{\alpha\beta \ \ \rho\sigma}_{\ \ \gamma\delta}(p,r)T^{\gamma\delta}(-r) h_{\alpha\beta}(p)  =  -\left(h^{\alpha\beta}(p)r_{\alpha}r_{\beta} - p_{(\alpha}r_{\beta)}h^{\alpha\beta}(p)+\frac{1}{2}p\cdot r h(p)\right)T^{\rho\sigma}(-r)  \\ 
 +\left((p^2 -2p\cdot r)h_{\lambda}^{\ (\rho|}(p) + p_{\lambda}(p_{\alpha}-2r_{\alpha})h^{\alpha (\rho|}(p)
  -p^{(\rho|}(p_{\alpha}-2r_{\alpha})h^{\alpha}_{\ \lambda}(p)\right)T^{|\sigma)\lambda}(-r). 
\end{multline}
Since $(-\nabla^2)^m T^{\rho\sigma}$ is still a contravariant tensor, we can similarly pull out the $O(h)$ part 
from:
\begin{equation}
(-\nabla^2)^{n}(p,r) T^{\rho\sigma}(-r) = \sum_{m=0}^{n-1} |p-r|^{2(n-1-m)} H^{\alpha\beta \ \ \rho\sigma}_{\ \ \gamma\delta}(p,r)\, |r|^{2m}\, T^{\gamma\delta}(-r) h_{\alpha\beta}(p)\,,
\end{equation}
Summing the geometric progression:
\begin{equation}
 \sum_{m=0}^{n-1} |p-r|^{2(n-1-m)}|r|^{2m} = \frac{(p-r)^{2n}-r^{2n}}{(p-r)^2 -r^2}\,,
\end{equation}
and using \eqref{kernel-taylor} we \new{find the form of $\KoL$ at the 1-point level to be}
\begin{equation}
\dot{\Delta}(-\nabla^2)(p,r) T^{\rho\sigma}(-r) = \frac{\dot{\Delta}\left(|p-r|^2\right) - \dot{\Delta}(r^2)}{|p-r|^2-r^2}H^{\alpha\beta \ \ \rho\sigma}_{\ \ \gamma\delta}(p,r)T^{\gamma\delta}(-r)\new{h_{\alpha\beta}(p)} \old{+ O(h^2)}\,.
\end{equation}
After expanding the overall kernel to the desired order in $h$, one can take functional derivatives in the usual way to obtain $n$-point functions.
We will not dwell on this further because we will only need the 0-point function of the kernel in the remainder of this paper.

\section{Transverse 2-point functions}\label{sec:transverse}

The 2-point Ward identities (\ref{Ward1to2}) and \eqref{Ward2ptCC}
tell us that the momentum-dependent part of the 2-point function is transverse.
Although we can 
obtain the unique form of the momentum-independent part through (\ref{Ward2ptCC}),
we cannot use the Ward identities alone to obtain the momentum-dependent part.
In this section, we demonstrate that there exist two linearly independent transverse 2-point structures that respect the required diffeomorphism invariance of the action.
Momentum conservation tells us that there is only a single momentum argument, $p$, at the 2-point level.
Structures that are of odd order in the momentum are forbidden by Lorentz invariance, so let us begin with quadratic structures.
The most general structure that is at quadratic order in the metric perturbation and momentum is
 \begin{equation}
 a_{1}h p^2 h + a_{2}h_{\alpha\beta}p^2 h^{\alpha\beta} + a_{3}hp_{\alpha}p_{\beta}h^{\alpha\beta} + a_{4}h^{\alpha\beta}p_{\alpha}p_{\gamma}h_{\beta}^{\ \gamma}\,,
\end{equation}
where the $a_i$ are numerical coefficients.
Performing a linearized diffeomorphism $\delta h_{\alpha\beta}\to 2p_{(\alpha}\xi_{\beta)}$ and requiring this to vanish gives
$a_{1}=-a_{2}=-a_{3}/2=a_{4}/2$. Thus we have only one allowed structure that is quadratic in the momentum:
\begin{equation}
\label{EH2}
\cL^{(2)}_{EH} =
  \frac{1}{2}\left(h_{\mu\nu}p^2 h^{\mu\nu}-hp^2 h+2h^{\mu\nu}p_{\mu}p_{\nu}h-2h^{\mu\nu}p_{\mu}p_{\rho}h_{\nu}^{\ \rho}\right) \,.
\end{equation}
This corresponds to the Einstein-Hilbert action since 
\be
\int_{x}\!\!\sqrt{g}\,\cO_2= -2\int_{x}\!\!\sqrt{g}R =  \int\! \dbar p \ 
\cL^{(2)}_{EH} +O(h^3)\,.
\ee
We have a more general structure with quartic terms in momenta:
\be
 b_{1}h^{\alpha\beta}p^4 h_{\alpha\beta} + b_{2}hp^4 h + b_{3}h^{\alpha\beta}p^2 p_{\alpha}p_{\beta}h 
 + b_{4}h^{\alpha\beta}p^2 p_{\alpha}p_{\gamma}h_{\beta}^{\ \gamma} + b_{5}h^{\alpha\beta}p_{\alpha}p_{\beta}p_{\gamma}p_{\delta}h^{\gamma\delta} \,.
\ee
Requiring this to vanish under linearised diffeomorphisms 
%
%
gives us $b_{5}= b_{1} + b_{2}$, $b_{4}=-2b_{1}$, $b_{3}=-2b_{2}$, and thus leaves only two linearly independent transverse structures: 
\begin{eqnarray}
\cL^{(2)}_{a} &=& \half 
\left(
h^{\mu\nu}p^4 h_{\mu\nu} - 2h^{\mu\nu}p^2 p_{\mu}p_{\rho}h_{\nu}^{\ \rho} + h^{\mu\nu}p_{\mu}p_{\nu}p_{\rho}p_{\sigma}h^{\rho\sigma}\right)\,,\\
\cL^{(2)}_{b} &=& \half 
\left(
hp^4 h -2h^{\mu\nu}p^2 p_{\mu}p_{\nu}h +h^{\mu\nu}p_{\mu}p_{\nu}p_{\rho}p_{\sigma}h^{\rho\sigma}\right)\,.
\end{eqnarray}
These are now the most general index structures for $O(h^2)$,
since higher orders in momentum would have to be contracted into $p^2$ factors. Therefore the general form of the transverse two-point vertex at $O(p^4)$ and higher is given by the linear combination $a\,\cL^{(2)}_{a}+b\,\cL^{(2)}_{b}$, where $a(p^2/\Lambda^2)$ and $b(p^2/\Lambda^2)$ are Taylor expandable functions.
On the other hand\new{,} the Einstein-Hilbert structure \eqref{EH2} is also reproduced by setting $a=-b=1/p^2$. 

The choice of $b=2$, $a=0$ gives the 2-point part of the $R^2$ term in the action.
Similarly $a=2$, $b=0$ gives the 2-point part of the $R_{\mu\nu\rho\sigma}R^{\mu\nu\rho\sigma}$ term.
Also, $a=b=1/2$ gives the 2-point part of the $R_{\mu\nu}R^{\mu\nu}$ term. (The linear relation between these three 2-point vertices is of course the one implied by the Gauss-Bonnet topological invariant.)

We can express these structures explictly as 2-point functions as follows:
\bea
\label{EH2pt}
\mathcal{S}^{\mu\nu\rho\sigma}_{\rm EH}(-p,p) &=& p^2(\delta^{\mu(\rho}\delta^{\sigma)\nu}-\delta^{\mu\nu}\delta^{\rho\sigma})+p^{\mu}p^{\nu }\delta^{\rho\sigma}+p^{\rho}p^{\sigma }\delta^{\mu\nu}-2p^{(\mu |}p^{(\rho}\delta^{\sigma)|\nu)}\,,\\
\label{a2pt}
\mathcal{S}^{\mu\nu\rho\sigma}_{a}(-p,p) &=& 
p^4\delta^{\mu(\rho}\delta^{\sigma)\nu}-2p^2 p^{(\mu |}p^{(\rho}\delta^{\sigma) |\nu)} +p^{\mu}p^{\nu}p^{\rho}p^{\sigma} \\
&=& \left(p^2 \delta^{(\mu |(\rho}-p^{(\mu |}p^{(\rho}\right)\left(p^2 \delta^{\sigma)|\nu)}-p^{\sigma)}p^{|\nu)}\right)\,,\nonumber\\
\label{b2pt}
\mathcal{S}^{\mu\nu\rho\sigma}_{b}(-p,p) &=& 
p^4\delta^{\mu\nu}\delta^{\rho\sigma}-p^2p^{\mu}p^{\nu}\delta^{\rho\sigma}-p^2p^{\rho}p^{\sigma}\delta^{\mu\nu}+p^{\mu}p^{\nu}p^{\rho}p^{\sigma} 
\\
&=& \left(p^2\delta^{\mu\nu}-p^{\mu}p^{\nu}\right)\left(p^2 \delta^{\rho\sigma}-p^{\rho}p^{\sigma}\right)\,.\nonumber
\eea


\section{Tree-level 2-point functions at fixed points}
\label{sec:2pts}

Since neither \new{choice of} fixed point and seed Lagrangian\new{,} \eqref{Shat-4} or \eqref{biEHfull}\new{,} involves the unit operator (\aka cosmological constant) term, the two-point vertex will be transverse for both schemes.
Using the linearly independent structures from the previous section, we can now derive their exact classical fixed point 2-point graviton vertices through the flow equation \eqref{a0}, thus
relating these consistently to the form of $\KoL$ and the cutoff profile $c$.

\subsection{Weyl scheme 2-point vertex}\label{sec:fbfixed}

As anticipated in sec. \ref{sec:bifixed}, we can set the seed Lagrangian to be \eqref{Shat-4}. Since we want the 2-point vertex of the classical fixed point action to coincide with this, we have for both actions that
\begin{equation} 
\label{fullaction2}
 \mathcal{S}^{\alpha\beta\gamma\delta} = c^{-1}\mathcal{S}^{\alpha\beta\gamma\delta}_{a}+\left(1+ 4s\right)c^{-1}\mathcal{S}_{b}^{\alpha\beta\gamma\delta}\,,
\end{equation}
using the notation for 2-point functions in (\ref{a2pt}) and (\ref{b2pt}). From \eqref{t.t.} we get the ``two-traces'' part of the flow
\begin{equation}\label{f2ptt}
  -16(1+3s)^2c^{-2}p^{4}\KoL\,\mathcal{S}^{\new{\alpha\beta\gamma\delta}}_{b}(p,-p)\,,
\end{equation}
and from \eqref{c.c.} the  ``cross-contracted'' part
\be\label{f2pcc}
   -4(1+2s)(1+6s)\,c^{-2}p^{4}\KoL\,\mathcal{S}^{\new{\alpha\beta\gamma\delta}}_{b}   - c^{-2} p^4 \KoL\left(\mathcal{S}^{\new{\alpha\beta\gamma\delta}}_{a}+\mathcal{S}^{\new{\alpha\beta\gamma\delta}}_{b}\right)\,.
\ee
and thus comparing \eqref{flowratio} to the RG-time derivative of (\ref{fullaction2}):
\bea\label{diff1}
 \dot{\left(c^{-1}\right)} &=& -p^4 c^{-2} \KoL\,,\\
\label{diff2}
 s\dot{\left(c^{-1}\right)} &=& -p^4 c^{-2} \KoL \left[4j(1+3s)^2+(1+2s)(1+6s)\right]\,.
\eea
Requiring \eqref{diff2} to be consistent with \eqref{diff1} determines $j= - (1+4s)/4(1+3s)$ \ie  the value \eqref{j-Weyl} determined in the background-independent calculation, while 
eqn. (\ref{diff1}) itself is solved by the normalised choice \eqref{prop-Weyl} already put forward for the effective propagator. 

\subsection{Einstein scheme 2-point vertex}\label{sec:fbEHfixed}

As anticipated in sec. \ref{sec:biEHfixed}, we will see that we can set the seed Lagrangian to be \eqref{biEHfull}. Since we want the 2-point vertex of the classical fixed point action to coincide with this, we have for both actions that
\begin{equation}
\label{fullEaction2}
 \mathcal{S}^{\alpha\beta\gamma\delta} = \left(\frac{1}{p^2}+\frac{d}{\Lambda^2}\right)\mathcal{S}^{\alpha\beta\gamma\delta}_{a}+\left(-\frac{1}{p^2} + \left(1 + 4j\right)\frac{d}{\Lambda^2}\right)\mathcal{S}_{b}^{\alpha\beta\gamma\delta}\,.
\end{equation}
From \eqref{t.t.} we get the ``two-traces'' part of the flow
\begin{equation}\label{fEHtt}
 -4p^{4}\left(\frac{1}{p^2}-2(1+3j)\frac{d}{\Lambda^2}\right)^2\!\KoL\,\mathcal{S}^{\new{\alpha\beta\gamma\delta}}_{b}\,,
\end{equation}
and from \eqref{c.c.} we get the ``cross-contracted'' part:
\be\label{fEHpcc}
4(1+2j)p^4\frac{d}{\Lambda^2}\left(\frac{2}{p^2}-(1+6j)\frac{d}{\Lambda^2}\right)
\KoL\,\mathcal{S}^{\new{\alpha\beta\gamma\delta}}_{b}  - p^4
\left(\frac{d}{\Lambda^2} +\frac{1}{p^2}\right)^2
 \!\KoL\left(\mathcal{S}^{\new{\alpha\beta\gamma\delta}}_{a}
+\mathcal{S}^{\new{\alpha\beta\gamma\delta}}_{b}\right)\,.
\ee
Thus comparing \eqref{flowratio} to the RG-time derivative of (\ref{fullEaction2}):
\bea 
\label{diffE1}
\Lambda\partial_{\Lambda}\left(\frac{d}{\Lambda^2}\right) &=& -p^4 \KoL\left(\frac{d}{\Lambda^2} +\frac{1}{p^2}\right)^2\,,\\
\label{diffE2}
 j\Lambda\partial_{\Lambda}\left(\frac{d}{\Lambda^2}\right) &=& -p^4 \KoL\left(\frac{j}{p^4}+(1+12j+36j^2+36j^3)\frac{d^2}{\Lambda^4}-2(1+4j+6j^2)\frac{d}{p^2 \Lambda^2}\right).
\eea
The first equation is solved by the assumed effective propagator \eqref{prop-Einstein}, providing we identify
\be 
c= \frac{1}{1+d\,p^2\!/\Lambda^2}\,.
\ee
First order expansion confirms the relation $c'(0)=d(0)$ we found from the background-independent calculation, \cf below \eqref{biEHfull}. On the other hand \eqref{diffE1} and \eqref{diffE2} are consistent if and only if $j=-1/2$ or $j=-1/3$. 
From \eqref{fullEaction2}, we see that the latter solution implies that the
index structure of the regularised 2-point vertex is not identical to the Einstein-Hilbert term. If we choose the $j=-1/2$ solution however the classical fixed point and seed-action 2-point vertex is simply
\be 
\mathcal{S}^{\alpha\beta\gamma\delta}(p,-p) = c^{-1} \mathcal{S}^{\alpha\beta\gamma\delta}_{\rm EH}\,.
\ee


\section{Discussion and Conclusions}
\label{sec:conclusions}

In this paper we have constructed a manifestly diffeomorphism invariant continuum Wilsonian RG (\aka exact RG) at the classical level (by which we mean precisely the $\hbar\to0$ limit \cf \eg the discussion in sec. \ref{sec:biEHfixed}), and sketched the first steps for quantum gravity. As addressed at the beginning of the Introduction, already the classical construction could be useful. Indeed, since gravity is very weakly coupled at currently accessible scales, the classical level applies to all currently observed gravitational physics.
The formulation allows computations to be done by phrasing the problem in terms of computing the Wilsonian effective action $S[g]$ at some diffeomorphism preserving effective momentum cutoff scale $\Lambda$. Although we have not discussed this here, the formulation allows in principle to compute exactly the expectation of any diffeomorphism invariant operator, along the lines of refs. \cite{Rosten:2006qx,Rosten:2006pd} for example. It is important to emphasise that the effective action $S$ is arrived at by an exact transformation from the original ``bare'' action. At the quantum level, this was demonstrated in general in sec. \ref{sec:kadanoff}. We  gave an independent demonstration of this for classical gravity in eqn. \eqref{classical_equivalence}. Therefore no information is actually lost by ``integrating out'' modes down to the effective cutoff $\Lambda$.

By utilising the freedom to design the Kadanoff blocking (see the review in sec. \ref{sec:review} and application to gravity in sec. \ref{sec:biflow}) it is actually straightforward to ensure that the flow equation respects diffeomorphism invariance. More surprising perhaps is the fact that the effective action can then be explicitly computed without gauge fixing. One way to do this is to start by following standard practice, and pick a space-time manifold and convenient coordinates, \eg flat, and perturb about a ``background'' metric \eg $\bar{g}_{\mu\nu} = \delta_{\mu\nu}$. The difference here is that no gauge fixing step is required and thus the (differential) Ward identities expressing exact diffeomorphism invariance, are obeyed. These confirm that the momentum-independent piece becomes the cosmological constant term, and that the remaining two-point vertex is transverse. They also show how to relate the ($n$+1)-point vertex with one zero momentum argument to an $n$-point vertex, thus closing the flow equations at the classical level and allowing the $n$-point vertices to be computed iteratively in terms of the lower point vertices.
We developed this approach in the latter half of the paper, secs. \ref{sec:fbflow} -- \ref{sec:2pts}.

However it is not necessary to introduce a background metric, nor particular coordinates, nor even to pick a particular space-time manifold in order to compute $S$. It is a fundamental requirement that the flow equation, and also the solution $S$, be quasi-local \ie have vertices that are Taylor expandable to all orders in momenta.\footnote{A different notion of locality for quantum gravity has recently been discussed in ref. \cite{Christiansen2015}.} This encodes the requirement that the Kadanoff blocking effectively operates only on a local patch of the manifold. Nevertheless it is important to recognise that the implementation does not require flat space, or to be somehow close to flat space, rather the size of the patch is controlled in a diffeomorphism invariant and background-independent way by $1/\Lambda$, through the cutoff function $c(-\nabla^2/\Lambda^2)$, where $\nabla$ is the full quantum covariant derivative. In practical terms\new{,} it means that $S$ can be computed in terms of the full metric $g_{\mu\nu}$ simply by manipulating covariant derivatives.  The computation proceeds iteratively as an expansion in local diffeomorphism invariant operators of increasing engineering dimension. We pursued this approach in secs. \ref{sec:biflow} and \ref{sec:bieffact}. Although we do not do so here, it would require only minor modifications to phrase the computation in this framework entirely in coordinate free language.

It should be clear that it is the same effective action that we are computing by either fixed background or background-independent methods. We do however confirm this in a number of examples\new{. I}n sec. \ref{sec:bifixed}\new{,} we demonstrate (by obtaining the same value of $j$) that\new{,} in the Weyl scheme \old{in the }background-independent computation\new{,} the same two-curvature $\nabla^4$ and $\nabla^6$ terms arise as in the fixed background computation in sec. \ref{sec:fbfixed}\new{. W}e derive the same behaviour of the differential Ward identity \eqref{diffWardcontracted} from the background-independent computation \eqref{0-1} as explained at the end of sec. \ref{sec:fbflow}\new{. Finally,} in the Einstein scheme\new{,} we demonstrate in secs. \ref{sec:biEHfixed}, \ref{sec:fbEHfixed} that the coefficients of the curvature-squared operators are the same in the two approaches.

As stated already, it is actually quite straightforward to incorporate exact diffeomorphism invariance. Essentially one replaces the kernel $\dot{\Delta}_{xy}$ as it appears in the scalar flow equation \eqref{scalarSdot} by some appropriate covariantization $\{\dot{\Delta}\}_{xy}$. There is a great deal of freedom \old{however} in this. Following the treatment in gauge theory \cite{Morris:1998kz,Morris:1999px,Morris:2000fs}, we could have kept this general. We could have represented this as a weighted functional integral over path ordered integrals between $x$ and $y$ using the connection $\Gamma^\mu_{\alpha\beta}$. Instead we made perhaps the simplest choice which was to express $\dot{\Delta}$ as a differential operator and replace the partial differentials by covariant derivatives.

As we emphasised, there still remains a great deal of freedom in designing the exact RG, equivalently in the choice of $\Psi$ in \eqref{generalERG}. However while any choice of $\Psi$ that is quasi-local generates a quasi-local exact reparametrisation of the theory, as sketched below \eqref{generalERG}, it is not true that any choice leads to a valid exact RG. The key extra property we look for in the latter is that momenta are indeed effectively cutoff by $\Lambda$. For a fully quantum exact RG we can expect that extra structure is required, beyond the covariant higher derivatives introduced here, just as it was for gauge theory \cite{Morris:1999px,Morris:2000fs,Morris:2000jj,Arnone:2000bv, Arnone:2000qd,Arnone:2001iy}. But even before we consider this extra structure, it is still not true that there is complete freedom in choice of $\Psi$. For example $\Psi$ must depend on the effective action itself, otherwise the flow is linear inhomogeneous in \eqref{generalERG2} and cannot lead to fixed point behaviour. A slightly less straightforward example is given by discarding the seed action, \ie setting $\hat{S}=0$. \new{In that case,} in \eqref{scalarSdot}, $\Sigma=S$, so the flow is non-linear and at first sight is a valid starting point. However as we see in app. \ref{app:seed0}, the tree-level corrections then do not take the right form for the momentum integrals in the quantum corrections to be properly regulated.

To avoid such dangers, we chose to mimic what has already proved to work well for scalar and gauge theory: the significant choice being to require that the two-point vertex of the seed action be equal to the two-point vertex of the fixed point effective action. This in turn determines the form of the kernel $\dot{\Delta}$. As we have seen we can then arrange for sensible intuitive results in the sense that $\Delta$ comes out  as might be expected for an effective propagator for graviton fluctuations, mimicking the successful construction for gauge theory. However note that  these requirements, which guide the construction of the exact RG, mean that there remains some association with a preferred background (here flat) and indeed preferred expansion \eqref{h}, in the sense that it is this expansion about such a background that defines the two-point vertices of the fixed point and seed actions, which are then required to coincide.

Even after making these choices, there is still freedom. In particular\new{,} we set up two different versions which we called the ``Einstein scheme'' (secs. \ref{sec:biEHfixed}, \ref{sec:fbEHfixed})
and the ``Weyl scheme'' (secs. \ref{sec:bifixed}, \ref{sec:fbfixed})\com{the fixed-background links got mixed up, I fixed them here}. The Einstein scheme gives a privileged  r\^ole to Newton's constant $G(\Lambda)$, as an expansion in this irrelevant coupling around the Gaussian fixed point, equivalently an expansion in $1/M^2(\Lambda)$ where $M$ is the running Planck mass. The Weyl scheme is also an expansion around the Gaussian fixed point, but adapted to four-derivative gravity, a renormalisable theory with asymptotically free couplings but which has issues with unitarity \cite{Stelle:1976gc,Adler:1982ri}. 

A further apparent freedom appears in the index structure for the kernel \eqref{kernel-grav}, where it is parametrised by $j$. While this \old{clearly} has the same origin as the DeWitt supermetric \cite{DeWitt:1967yk} we find that for both schemes it is actually determined by the other choices we make (and for the Weyl scheme also by the fixed-point ratio $s(\omega_*)$ of the couplings). We only touched briefly on the special values $j=\infty$ and $j=-1/D$, which correspond to the conformal truncation \cite{Machado:2009ph,Reuter:2008qx,Reuter:2008wj,Bonanno:2012dg,Dietz:2015owa} and unimodular gravity \cite{Einstein1919,Unruh1989a,Eichhorn:2015bna,Saltas:2014cta} respectively. Since these variants can thus be naturally incorporated, it would be very interesting to develop them further.

As we noted in the introduction,  quantum corrections are not yet sufficiently regulated.
These are generated by
the second term in \eqref{full-flow}. If it is treated perturbatively, using the expansion around \new{a} fixed background\new{,} developed in the second half of this paper, we would find that the loop integrals suffer ultraviolet divergences. The problem that has to be faced is that the diffeomorphism invariant cutoff function $c(-\nabla^2/\Lambda^2)$, which is effectively covariant higher derivative regularisation, is not sufficient to regulate all ultra-violet divergences. One loop divergences slip through just as they do for gauge theories \cite{Slavnov:1972sq,Lee:1972fj}. Therefore extra ultraviolet regularisation needs to be incorporated into the exact RG flow equation. 

As we briefly reviewed in the introduction, for a gauge theory this extra regularisation is provided by generalising the gauge group from $SU(N)$ to $SU(N|N)$ and then spontaneously breaking the fermionic gauge fields at the effective cutoff scale $\Lambda$. The resulting massive fields behave as gauge invariant Pauli-Villars fields with masses set by $\Lambda$ and interactions that are naturally incorporated into the flow equation \cite{Morris:1998kz,Morris:1999px,Morris:2000fs,Morris:2000jj}. The reason these provide the needed extra regularisation can be understood as follows. The extra structure introduces as many wrong-statistics fermionic fields as there are bosonic degrees of freedom.\footnote{Actually for the counting to work exactly at finite $N$, it is first necessary to extend the group to $U(N|N)$ after which one sees that two vector bosons decouple \cite{Arnone:2001iy}.} 
For the gauge fields themselves, the original gauge field $A^1_\mu$ is joined by a copy gauge field $A^2_\mu$ and complex pair of fermionic gauge fields $B_\mu, \bar{B}_\mu$.
At high energies these degrees of freedom cancel each other, as happens with Parisi-Sourlas supersymmetry \cite{Parisi:1979ka}, at least sufficiently that\new{,} together with appropriately chosen covariant cutoff functions\new{,} the theory is then regularised to all orders in perturbation theory \cite{Arnone:2000bv, Arnone:2000qd,Arnone:2001iy}.

Given the developments just described it is natural to conjecture that the extra regularisation for gravity can be incorporated by introducing wrong-statistics fermionic components to the metric in a way that extends the diffeomorphism invariance along fermionic directions. We are therefore led naturally to consider extending the coordinates themselves to
\be 
x^A = (x^\mu,\theta^a)\,,
\ee
where, in Euclidean signature, the $D$ dimensional bosonic coordinates run from $\mu =1,\cdots,D$, while an equal number of real fermionic coordinates run from $a=D+1,\cdots,2D$. Writing the invariant interval as
\be 
ds^2 = dx^A g_{AB} dx^B\,,
\ee
we have introduced $D^2$ wrong-statistics fermionic degrees of freedom $g_{\mu a}=-g_{a\mu}$\new{,} which is the right number to cancel the $D^2$ bosonic degrees freedom, namely the $D(D+1)/2$ degrees of freedom in the original metric $g_{\mu\nu}$ and the $D(D-1)/2$ bosonic degrees of freedom in the antisymmetric components $g_{ab}$. We see that we are led to construct a particular type of supermanifold, what we might call a Parisi-Sourlas supermanifold. Fortunately\new{,} supermanifolds in general have been extensively developed \cite{DeWitt:1992cy}. 

Of course it remains to demonstrate whether this structure can indeed provide the missing regularisation and then also how to decouple the extra degrees of freedom at energies lower than $\Lambda$. Again\new{,} following the hints from gauge theory\new{,} we would expect to incorporate a running spontaneous symmetry breaking.
Possible strategies for the latter would be to consider extra fields, or particular structures in the Lagrangian or maybe even just particular solutions for $g_{AB}$.

\section*{Acknowledgements}

TRM acknowledges support from STFC through Consolidated Grant ST/L000296/1. AWHP acknowledges support from the University of Southampton through a Mayflower scholarship.

\appendix

\section{Expansion of the metric determinant}\label{sec:sqrtg}

The momentum-independent part of the action has the same $n$-point structure as $\sqrt{g}$, meaning that it corresponds to a cosmological constant-like term.
All actions carry a factor of $\sqrt{g}$, whereas the kernel carries a factor of $1/\sqrt{g}$.
Here, we list the first few $n$-point functions from  the $l$th power:
\begin{equation}
{\det{}^{l/2}(g_{\mu\nu})} = e^{\frac{l}{2}\tr\left(\ln\left(\delta_{\mu\nu}+h_{\mu\nu}\right)\right)}\,.
\end{equation}
Expanding out the logarithm gives the trace as 
$h - \frac{1}{2}h_{\mu\nu}h^{\mu\nu} + \frac{1}{3}h_{\mu\nu}h^{\mu\rho}h^{\nu}_{\ \rho} - \cdots$.
Then we expand the exponential to get
\begin{equation}
\sqrt{g}^{\,l} = 1 + l\frac{h}{2}-l\frac{h_{\mu\nu}h^{\mu\nu}}{4}+l^2\frac{h^2}{8}+l\frac{h_{\mu\nu}h^{\mu\rho}h^{\nu}_{\ \rho}}{6}-l^2\frac{h_{\mu\nu}h^{\mu\nu}h}{8}+l^3\frac{h^3}{48}+\cdots
\end{equation}
We can then obtain $n$-point functions by differentiating with respect to metric perturbations:
\begin{equation}
\label{one-point-cc}
\mathcal{S}_{c}^{\mu\nu} = \frac{l}{2}\delta^{\mu\nu},
\end{equation}
\begin{equation}
\label{two-point-cc}
\mathcal{S}_{c}^{\mu\nu\rho\sigma} = \frac{l^2}{4}\delta^{\mu\nu}\delta^{\rho\sigma} - \frac{l}{2}\delta^{\mu(\rho}\delta^{\sigma)\nu},
\end{equation}
\begin{eqnarray}
\label{three-point-cc}
{\mathcal{S}}_{c}^{\mu\nu\rho\sigma\alpha\beta} & = & \frac{l^3}{8}\delta^{\mu\nu}\delta^{\rho\sigma}\delta^{\alpha\beta}
+l\delta^{(\mu|(\rho}\delta^{\sigma)(\alpha}\delta^{\beta)|\nu)} \nonumber \\ &&
-\frac{l^2}{4}\left(\delta^{\mu\nu}\delta^{\rho(\alpha}\delta^{\beta)\sigma}
+\delta^{\rho\sigma}\delta^{\mu(\alpha}\delta^{\beta)\nu}
+\delta^{\alpha\beta}\delta^{\mu(\rho}\delta^{\sigma)\nu}\right).
\end{eqnarray}
The choice of $l=1$ gives the $n$-point functions implied by the momentum-independent Ward identities in (\ref{Wardmomind}), as seen by explictly comparing (\ref{Ward2ptCC}) and (\ref{Ward3ptCC}) to (\ref{one-point-cc}), (\ref{two-point-cc}) and (\ref{three-point-cc}).

\section{Why the seed action cannot be set to zero}\label{app:seed0}

We demonstrate that the choice $\hat{S}=0$ does not lead to an acceptable exact RG. For this purpose we can work with the $\varphi\leftrightarrow-\varphi$ invariant scalar field theory treated in sec. \ref{sec:scalars}. Setting $\hat{S}=0$ in \eqref{scalarSdot} means that the classical flow equation is simply
\be 
\label{seed0-flow}
\dot{S} = \frac{1}{2} \frac{\delta S}{\delta\varphi}\cdot\dot{\Delta}\cdot\frac{\delta S}{\delta\varphi}\,.
\ee
Then instead of \eqref{scalar-2pt-flow} we have 
\be 
\label{seed0-flow-2pt}
\dot{S}^{(2)} = \dot{\Delta} \left(S^{(2)}\right)^2\,.
\ee 
(In this appendix we will use $S^{(n)}(p_1,\cdots,p_n)$ to denote the effective action $n$-point vertex with the momentum conserving $\delta$-function factored out.) Thus we are now led to the choice $\Delta=-1/S^{(2)}$. From \eqref{seed0-flow}, the four-point vertex satisfies the flow equation:
\be 
\label{seed0-flow-4pt}
\dot{S}^{(4)}(p_1,\cdots,p_4)=S^{(4)}(p_1,\cdots,p_4)\,\old{\half}\sum_{i=1}^4 \dot{\Delta}(p_i) S^{(2)}(p_i)\,,
\ee
where $S^{(2)}(p)$ is short hand for $S^{(2)}(p,-p)$. Using \eqref{seed0-flow-2pt} we see that this has solution:
\be 
S^{(4)}(p_1,\cdots,p_4) = S^{(4)}_0(p_1,\cdots,p_4)\, \prod_{i=1}^4 \new{S^{(2)}(p_i)}\,,
\ee
where the integration `constant' is a $\Lambda$-independent Taylor-expandable four-point vertex $S^{(4)}_0$. (Standard RG considerations would lead us to set this simply to a four-point coupling $\lambda$.) The two-point vertex decoration shown in \eqref{seed0-flow-4pt} appears for any $n$-point vertex, for example the six-point vertex flow takes the form:
\begin{multline}
\dot{S}^{(6)}(p_1,\cdots,p_6)=S^{(6)}(p_1,\cdots,p_6)\,\old{\half}\sum_{i=1}^6 \dot{\Delta}(p_i) S^{(2)}(p_i)\\
+ \new{\half}\sum_{\rm partitions\ \pi}\!\!\! S^{(4)}(p_{\pi_1},p_{\pi_2},p_{\pi_3},-P)\dot{\Delta}(P) S^{(4)}(P,p_{\pi_4},p_{\pi_5},p_{\pi_6})\,,
\end{multline}
where $P=p_{\pi_1}+p_{\pi_2}+p_{\pi_3}$. Thus all tree-level interactions vertices have $\new{S^{(2)}}$ on their external legs, as in \eqref{seed0-flow-4pt}, where they appear as integrating factors. Loop corrections follow from the second term in \eqref{scalarSdot}. We see that the propagator in the loop thus appears with the factors:
\be 
\new{S^{(2)}} \Delta \new{S^{(2)}} =\new{S^{(2)}}\,,
\ee
which has the incorrect momentum dependence\new{, since it takes the form of a 2-point function rather than a UV regularized propagator, which would be its inverse}.

\bibliographystyle{hunsrt}
\bibliography{references}

\end{document}